\documentclass[sigconf, nonacm]{acmart}

\usepackage{soul}
\usepackage{algorithm}
\usepackage{algpseudocode}
\usepackage{caption}
\usepackage{subcaption}
\usepackage{booktabs}
\usepackage{multirow}
\usepackage{amsmath}
\usepackage{tikz}
\usepackage{xcolor}
\usepackage{xspace}
\usepackage{cleveref}
\usepackage{multicol}
\usepackage{balance}

\newcommand\vldbdoi{10.14778/3551793.3551844}
\newcommand\vldbpages{2953 - 2965}
\newcommand\vldbvolume{15}
\newcommand\vldbissue{11}
\newcommand\vldbyear{2022}
\newcommand\vldbauthors{\authors}
\newcommand\vldbtitle{\shorttitle} 
\newcommand\vldbavailabilityurl{https://github.com/uw-mad-dash/llamatune}
\newcommand\vldbpagestyle{empty} 

\usepackage[normalem]{ulem}

\newcommand{\revision}[1]{{#1}}
\newcommand{\revisionenv}[0]{}

\newcommand*\circled[1]{\tikz[baseline=(char.base)]{
            \node[shape=circle,fill,inner sep=1pt] (char) {\textcolor{white}{#1}};}}
\DeclareMathOperator*{\argmax}{argmax}

\newcommand{\sys}{LlamaTune\xspace}

\newcommand{\code}[1]{\texttt{#1}}

\begin{document}
\title{\sys: Sample-Efficient DBMS Configuration Tuning}


\author{Konstantinos Kanellis}
\affiliation{
    \institution{University of Wisconsin-Madison}
}
\email{kkanellis@cs.wisc.edu}

\author{Cong Ding}
\affiliation{
    \institution{University of Wisconsin-Madison}
}
\email{congding@cs.wisc.edu}

\author{Brian Kroth}
\affiliation{
    \institution{Microsoft Gray Systems Lab}
}
\email{bpkroth@microsoft.com}

\author{Andreas M{\"u}ller}
\affiliation{
    \institution{Microsoft Gray Systems Lab}
}
\email{amueller@microsoft.com}

\author{Carlo Curino}
\affiliation{
    \institution{Microsoft Gray Systems Lab}
}
\email{ccurino@microsoft.com}

\author{Shivaram Venkataraman}
\affiliation{
    \institution{University of Wisconsin-Madison}
}
\email{shivaram@cs.wisc.edu}

\begin{abstract}

Tuning a database system to achieve optimal performance on a given workload is a long-standing problem in the database community. A number of recent works have leveraged ML-based approaches to guide the sampling of large parameter spaces (hundreds of tuning knobs) in search for high performance configurations.
Looking at Microsoft production services operating millions of databases, \emph{sample efficiency emerged as a crucial requirement} to use tuners on diverse workloads. 

This motivates our investigation in \sys, a tuner design that leverages domain knowledge to improve the sample efficiency of existing optimizers.
\sys employs an automated dimensionality reduction technique based on randomized projections, a biased-sampling approach to handle special values for certain knobs, and knob values bucketization, to reduce the size of the search space.
\sys{} compares favorably with the state-of-the-art optimizers across a diverse set of workloads.
It identifies the best performing configurations with up to $11\times$ fewer workload runs, and reaching up to $21\%$ higher throughput.
We also show that benefits from \sys{} generalize across both BO-based and RL-based optimizers, as well as different DBMS versions.

While the journey to perform database tuning at cloud-scale remains long, \sys{} goes a long way in making automatic DBMS tuning practical at scale.

\end{abstract}


\maketitle

\pagestyle{\vldbpagestyle}
\begingroup\small\noindent\raggedright\textbf{PVLDB Reference Format:}\\
\vldbauthors. \vldbtitle. PVLDB, \vldbvolume(\vldbissue): \vldbpages, \vldbyear.\\
\href{https://doi.org/\vldbdoi}{doi:\vldbdoi}
\endgroup
\begingroup
\renewcommand\thefootnote{}\footnote{\noindent
This work is licensed under the Creative Commons BY-NC-ND 4.0 International License. Visit \url{https://creativecommons.org/licenses/by-nc-nd/4.0/} to view a copy of this license. For any use beyond those covered by this license, obtain permission by emailing \href{mailto:info@vldb.org}{info@vldb.org}. Copyright is held by the owner/author(s). Publication rights licensed to the VLDB Endowment. \\
\raggedright Proceedings of the VLDB Endowment, Vol. \vldbvolume, No. \vldbissue\ %
ISSN 2150-8097. \\
\href{https://doi.org/\vldbdoi}{doi:\vldbdoi} \\
}\addtocounter{footnote}{-1}\endgroup

\ifdefempty{\vldbavailabilityurl}{}{
\vspace{.3cm}
\begingroup\small\noindent\raggedright\textbf{PVLDB Artifact Availability:}\\
The source code, data, and/or other artifacts have been made available at \url{\vldbavailabilityurl}.
\endgroup
}

\section{Introduction}
\label{sec:intro}

Tuning the configuration of a database management system (DBMS) is necessary to achieve high performance. While DBMS tuning has been traditionally performed manually, e.g., by DB administrators, the shift to cloud computing, the increasing diversity of workloads and the large number of configuration knobs~\cite{storm2006adaptive, ituned} makes it challenging to manually tune databases for high performance. As a result, a number of automated methods~\cite{ottertune, ituned, cereda2021cgptuner, restune, rltuning, qtune, adaptive_rltuning,bestconfig, wang2021udo} for tuning have been proposed, with recent methods using machine learning (ML). ML-based methods have been shown to generalize to a number of workloads~\cite{zhang2021facilitating} and can find configurations that achieve up to 3-6$\times$ higher throughput (or lower latency)~\cite{qtune, ottertune, rltuning} when compared to using manually-tuned configurations.

\begin{figure}[tp]
    \centering
    \includegraphics[width=0.97\linewidth]{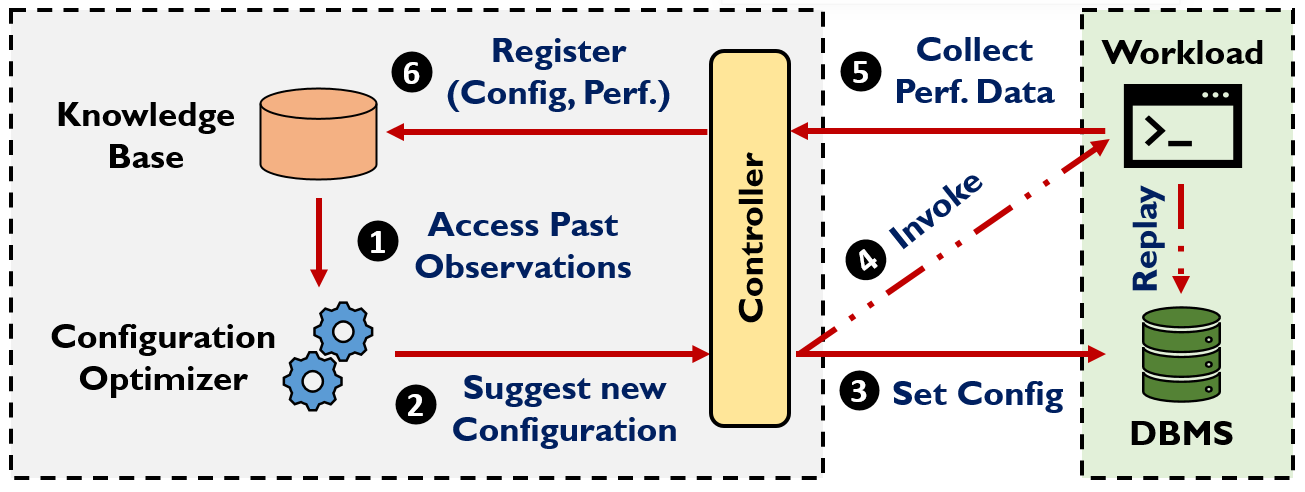}
    \vspace{-2ex}
    \caption{Overview of DBMS Knob Tuning Procedure}
    \label{fig:tuning_process}
    \vspace{-2ex}
\end{figure}

There are two main classes of ML-based approaches: those that perform prior training on selected benchmarks and transfer (or fine-tune) this knowledge given new customer workloads (e.g., OtterTune~\cite{ottertune} and CDBTune~\cite{rltuning}), and those \cite{qtune,ituned,cereda2021cgptuner} that directly tune a new customer workload by iteratively selecting a configuration using an optimizer, and running workloads with them (Figure~\ref{fig:tuning_process}). 
Algorithms used in these configuration optimizers try to balance between \emph{exploring} unseen regions of the configuration search space, and \emph{exploiting} the knowledge gathered until that point. Search-based methods~\cite{bestconfig}, RL-based methods~\cite{qtune} or Bayesian Optimization (BO) methods~\cite{cereda2021cgptuner, ituned} are commonly used, though a recent study~\cite{zhang2021facilitating} found that SMAC~\cite{hutter2011smac}, a BO-based method, performed best for DBMS tuning. 

In both settings, the cost and time involved in exploring several configurations before reaching a chosen one is the core challenge for practical relevance. Based on our visibility into production services operating millions of databases in Azure, we find that \emph{sample efficiency} (i.e., how few workload runs does it take to reach a given level of performance) is crucial to scale autotuning across diverse workloads.
Currently, state-of-the-art methods require 100 or more iterations~\cite{zhang2021facilitating} to converge to a good configuration, with each iteration taking several minutes\footnote{For certain workloads we might need to run each test much longer to saturate the system at a given VM size for instance~\cite{oltpbench}.}. To this purpose we focus our investigation in developing techniques that can improve sample efficiency, and that are broadly complementary to existing work.

We observe that existing optimizers do not leverage expert knowledge about the system they are tuning and using \textit{domain knowledge} has the potential to significantly improve sample efficiency.
First, based on our prior work~\cite{kanellis2020too} (and others~\cite{van2021inquiry, zhang2021facilitating}) we find that tuning a few important knobs is sufficient for achieving high DBMS performance, and tuning a smaller configuration space can lead to significant improvements in the number of samples required~\cite{zhang2021facilitating}. Yet, which knobs are important varies by workload, and existing methods for identifying important knobs are expensive and can be unreliable (Section~\ref{subsec:important-knobs}).
Second, we observe that not all DBMS knob values are the same.
For example, in PostgreSQL, the flag \code{backend\_flush\_after} denotes the number of pages after which previously performed writes are flushed to disk.
However, the flag has a special meaning when set to 0, disabling forced writeback and letting the OS manage it.
Such special values can affect the system in unpredictable ways and existing optimizers are unaware of such behaviors.
Additionally, we note that some parameters may have very large valid ranges. For instance, \code{shared\_buffers} is given as the number of 8K pages to account for potentially 100s of GBs.
Yet, the effect on performance between two nearby values, (e.g., 100 vs. 101), may be negligible.
These insights lead us to ask: \emph{How can we build a domain knowledge-aware configuration tuner that can minimize the number of samples required without sacrificing the final DBMS performance achieved?}

{\bf Contributions}
We address these challenges in \sys{} with the following contributions.
First, we show that surprisingly it is possible to achieve the benefits of only tuning important knobs but without using \textit{any} prior knowledge. Specifically, we leverage a randomized low-dimensional projection approach~\cite{hesbo} where we perform a projection of the configuration space from all dimensions ($D$) to a lower-dimensional subspace ($d$) and then use the optimizer to tune this smaller subspace. We also discuss how such a method can be integrated with a DBMS and handle categorical and numerical knobs.
Second, to handle knobs that have a different behavior for special values, we design a new biased-sampling based approach.
This makes the optimizer aware of the special case behavior with high probability during its random initialization phase and minimizes the number of samples required to find good configurations. 
To address large value ranges, we experiment with \emph{bucketizing} nearby values into groups (e.g., 100MB increments) to further reduce the search space that needs to be explored.
We find that this can improve convergence without sacrificing much from the performance of the optimal configuration for some workloads.

We combine the above techniques to design \sys \footnote{Llamas have been known to learn simple tasks with few repetitions~\cite{llama}.}, an end-to-end tuning framework that can tune a DBMS for new workloads without using any prior knowledge.
\revision{We integrate \sys{} with three state-of-the-art optimizers: two BO-based ones (SMAC~\cite{hutter2011smac}, GP-BO~\cite{ru2020bayesian}), and an RL-based one (DDPG~\cite{rltuning}).}
We evaluate \sys using six popular \revision{OLTP} workloads and consider scenarios where we tune for optimizing throughput or tail latency (95th percentile). 
Our results show that \sys can converge to the optimal configuration up to $11\times$ faster than SMAC while also reaching up to $20.85\%$ higher throughput.
\revision{We also demonstrate that similar gains can be achieved when \sys is used with GP-BO, or DDPG}. 
\revision{Finally, we also detail the process and our experiences of porting \sys to a newer DBMS version, and show that \sys can deliver meaningful gains with little effort}.

\section{Preliminaries}

In this section, we formalize the DBMS knob configuration problem and discuss existing methods for tuning.
Then, we identify some limitations of existing tuners and opportunities for improvements. 

\subsection{Background and Related Work}
\label{subsec:automatic-tuning}

\subsubsection{Automatic Database Tuning}
We can formulate database knob configuration tuning as an optimization problem.
Consider as input, a set of set of $n$ knobs, $\theta_1, \ldots, \theta_n$, alongside their respective domains $\Theta_1, \ldots, \Theta_n$.
These knobs domains can be either continuous, discrete, or categorical. 
Given this set of knobs, the resulting \textit{configuration space} can be defined as $\Theta = \Theta_1 \times \ldots \times \Theta_n$.
We assume a performance metric is encoded by the \textit{objective function} $f : \Theta \to \mathbb{R}$, that assigns a single performance value to each configuration.
Common performance metrics used in database tuning include system throughput, or tail latency (e.g., $95$th percentile latency).
Retrieving the value $f$ at some configuration point $\theta$ requires measuring this performance metric when the system is executed with $\theta$.
Given a workload, the goal of database tuning is to find a configuration $\theta^*$ that maximizes (or minimizes) the target metric:
$\theta^* = \argmax_{\theta \in \Theta} f(\theta)$

Most database knob tuning frameworks are built around a common architecture and
the main components include \emph{(i)} a knowledge base (KB), \emph{(ii)} a configuration optimizer, and \emph{(iii)} an experiment controller.
The knowledge base holds a record of all previously evaluated samples, $\mathcal{D} = \{\theta^j, f(\theta^j)\}$, and gets updated every time a new configuration is evaluated.
The configuration optimizer leverages the knowledge base to recommend configuration points $\theta^j$ that are promising (i.e., may lead to higher performance).
Finally, given a configuration, the controller orchestrates the execution of the workload and propagating the results (i.e., $\theta^j, f(\theta^j)$) back to KB.

Assuming a fixed workload, the tuning process is iterative as shown in~\autoref{fig:tuning_process}.
At first, \circled{1} the optimizer uses the current knowledge base, in order to \circled{2} suggest a configuration that it expects to improve performance.
Then, this configuration \circled{3} is applied to the DBMS instance running on a separate testing environment (green-shaded area).
Subsequently, \circled{4} the controller invokes the workload that feeds queries to the DBMS; the workload usually runs for several (e.g., $5$-$10$) minutes.
Once the workload execution completes, \circled{5} the controller collects the performance data (e.g., throughput) and any related metrics (e.g., internal DBMS metrics), and \circled{6} forwards them to the knowledge base, which is then updated with the observation (i.e., configuration, measured performance pair).
The above process is repeated until a certain budget $T$ (e.g., number of iterations, time, or cost limit) is exhausted.

Based on how much prior knowledge is available at the start of the tuning process, one can categorize various auto-tuners proposed in prior literature.
Some tuning frameworks~\cite{ottertune, rltuning, qtune, ituned} assume that before the tuning process begins, the KB has already been populated with prior observations.
In particular, these frameworks rely upon an initial profiling (or training) phase that includes evaluating thousands configuration points using benchmark workloads (e.g., up to $30$K for OtterTune~\cite{ottertune}).
This way, they can leverage prior knowledge gained from the training phase in order to propose good-performing configurations for new workloads. 
Other tuning frameworks~\cite{bestconfig, cereda2021cgptuner, wang2021udo} begin the tuning process without any prior knowledge: i.e., the KB is initially \emph{empty}.
Starting from zero knowledge, these tuners can only use the aforementioned tuning process to gather knowledge about the workload or the DBMS.
To combat this challenge, these tuners typically use algorithms that try to balance between \textit{exploring} unseen regions of the configuration search space, and \textit{exploiting} the knowledge gathered so far.
As the tuning process progresses and the KB accumulates more samples, these frameworks are eventually able to find configurations that achieve good performance.

In the above architecture, we can see that the configuration optimizer used plays a key role in determining the time (and hence cost) it takes to find a good performing configuration. Thus, we next survey configuration optimizers used in prior work.

\vspace{-1ex}
\subsection{Configuration Optimizers}
\label{subsec:existing-optimizers}
The configuration optimizer methods used by prior works can be classified into three categories: \textit{(i)} Search-based, \textit{(ii)} Reinforcement Learning (RL) based, and \textit{(iii)} Bayesian Optimization (BO) based.

A popular example of a search-based method is BestConfig~\cite{bestconfig}, which employs heuristics to divide the search space into multiple segments and navigates it through branch and bound policies.
Even though it can quickly choose which points to evaluate next, it does not makes use of a KB, which has been shown to hurt its performance~\cite{cereda2021cgptuner}.
Reinforcement learning methods proposed in \cite{rltuning, qtune, adaptive_rltuning} leverage the Deep Deterministic Policy Gradient (DDPG) algorithm~\cite{ddpg}, which trains an RL actor using a trial-and-error approach.
The actor utilizes a deep neural network that is trained based on DBMS internal metrics, and can propose configurations that balance exploration and exploitation.
In general, for RL methods to work well, thousands of samples have to be evaluated~\cite{qtune} and prior work~\cite{zhang2021facilitating} has found that RL based methods require more iterations due to the complexity of the neural networks used.

The majority of prior database auto-tuners use BO-based methods~\cite{ottertune, ituned, cereda2021cgptuner, restune, boat_paper}.
Bayesian optimization is a gradient-free, sequential, model-based approach that aims to find the optimum of an expensive-to-evaluate unknown function $f$, by evaluating as few samples as possible~\cite{shahriari2015taking}.
It also handles cases where the evaluated samples $f(\theta^j)$ are noisy.
BO consists of two main components: the \textit{surrogate model}, and the \textit{acquisition function}.
A surrogate model is an ML model, which given a set of observations $f(\theta^k)$, constructs an approximation of $f$; this approximation is refined each time a new point is evaluated.
The acquisition function is used to choose which point to evaluate next.
To do so, it uses the surrogate model to compute the expected utility of many candidate points, and then selects the one with the highest utility.
Utility values are calculated such that BO can effectively trade-off between the exploration of unseen regions, and the exploitation of good-performing ones.

While there are many BO variants, database knob tuners~\cite{ottertune, ituned, restune} have typically used ``vanilla'' Gaussian process (GP) as their surrogate model.
However, a recent comprehensive evaluation~\cite{zhang2021facilitating} of different BO variants shows that GP is not the optimal choice for database knob configuration tuning.
After experimenting with many modern black-box optimizers (including the ones used for database tuning~\cite{ottertune, ituned, cereda2021cgptuner, restune}) this study~\cite{zhang2021facilitating} concluded that the best performing optimizer was SMAC (i.e., Sequential Model-based Algorithm Configuration)~\cite{hutter2011smac}.
SMAC uses a random forest (RF) surrogate model, and is shown to perform very well in high-dimensional and heterogeneous search spaces, consisting of continuous, discrete, and categorical knobs.
Compared to vanilla GPs, RFs can inherently support categorical features, which helps find better configurations while also converging faster~\cite{hutter2011smac}.
Another method that improves over vanilla GP uses two separate kernels, Mat\'ern and Hamming, for continuous and categorical features, respectively.
This GP variant, which we refer to as GP-BO~\cite{ru2020bayesian}, also showed promising performance when dealing with a mixed search space, and a moderate number (i.e., up to $20$) of dimensions~\cite{zhang2021facilitating}.

\revision{In this work we focus on designing a domain knowledge-aware configuration optimizer and show that our techniques can be integrated with SMAC, GP-BO, and DDPG.}.

\subsection{Motivation}
\label{subsec:important-knobs}

\subsubsection{Only Tuning Important Knobs}
While database tuners that start with zero knowledge are able to suggest good-performing configurations eventually, they typically require many iterations until they can do so.
For instance, using the current state-of-the-art BO-variant, SMAC~\cite{hutter2011smac}, still requires around $100$ samples to reach optimal performance for most workloads~\cite{zhang2021facilitating}. With each sample evaluation (or iteration) taking $5-10$ minutes, the overall tuning process takes 8-16 hours. 
The main reason optimizers need a large number of iterations is that the configuration search space exposed to them is very \emph{high-dimensional}. 
The size of these search spaces are typically proportional to the number of configuration knobs that the user wishes to tune and this can include hundreds of knobs for popular databases like Postgres or MySQL~\cite{ottertune}.

\begin{table}[tp]
\centering
\small
\captionsetup{width=.48\textwidth}
\caption{SHAP's top-8 knobs vs hand-picked ones for \mbox{YCSB-A.} Underlined knobs indicate the differences between the two.}
\vspace{-1em}
\label{tab:top_8_shap_vs_manual}
\scalebox{0.97}{
\begin{tabular}{ll}
    \toprule
    \textbf{SHAP (top-8)} &\textbf{Hand-picked (top-8)} \\\midrule
    \ul{autovacuum\_vacuum\_threshold} &\ul{autovacuum\_analyze\_scale\_factor} \\
    autovacuum\_vacuum\_scale\_factor &autovacuum\_vacuum\_scale\_factor \\
    commit\_delay &commit\_delay \\
    \ul{enable\_seqscan} &full\_page\_writes \\
    full\_page\_writes &geqo\_selection\_bias \\
    geqo\_selection\_bias &\ul{max\_wal\_size} \\
    shared\_buffers &shared\_buffers \\
    wal\_writer\_flush\_after &wal\_writer\_flush\_after \\
    \bottomrule
\end{tabular}
}
\end{table}

\begin{figure}[tp]
    \centering
    \begin{subfigure}{0.46\linewidth}
        \includegraphics[width=\linewidth]{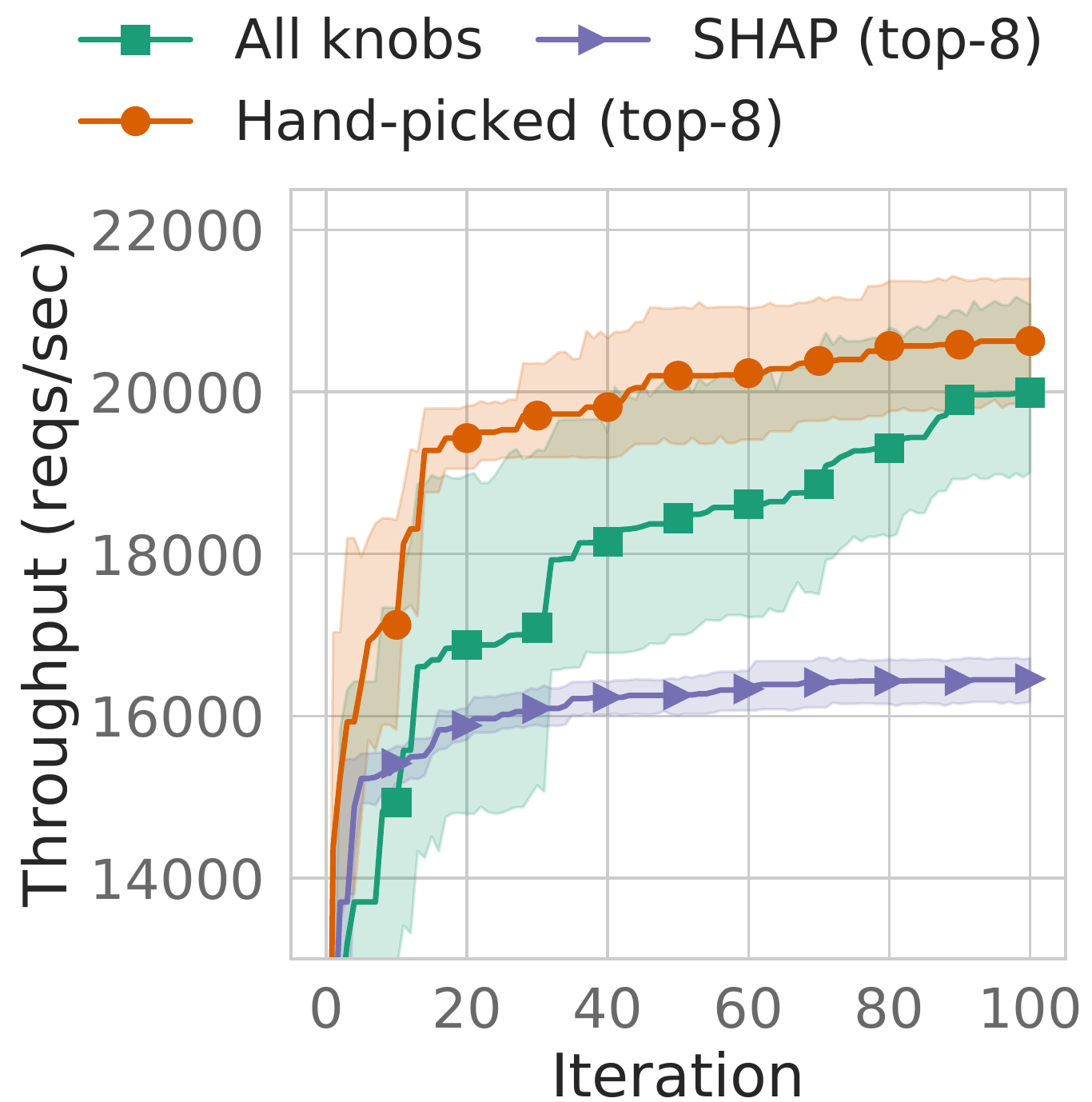}
        \vspace{-3ex}
        \caption{SHAP vs Manual (YCSB-A)}
        \label{fig:top_8_smac}
    \end{subfigure}
    \begin{subfigure}{0.48\linewidth}
        \includegraphics[width=\linewidth]{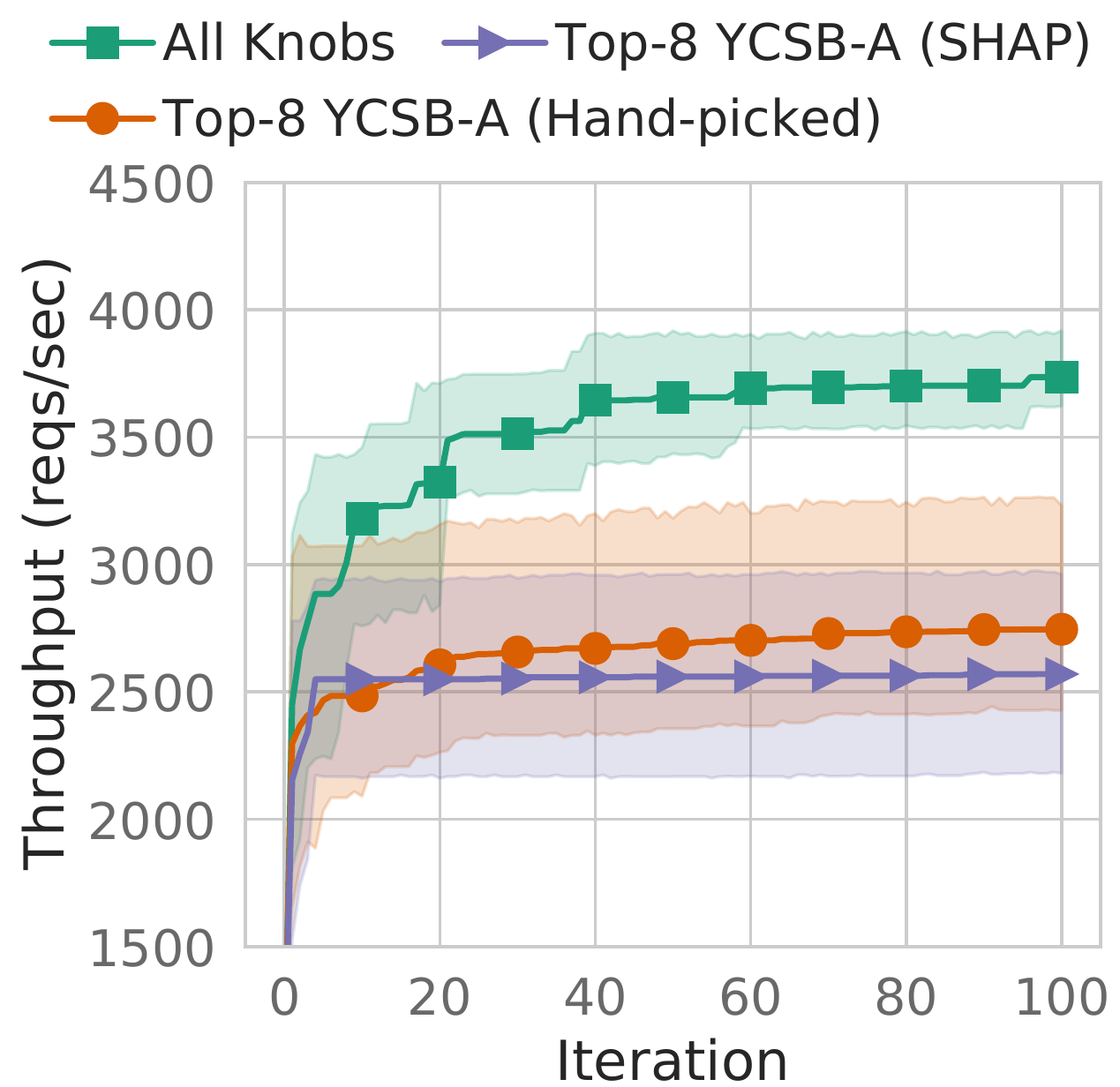}
        \vspace{-3ex}
        \caption{Top-8 YCSB-A to TPC-C}
        \label{fig:tpcc_top_8_ycsbA}
    \end{subfigure}
    \vspace{-2ex}
    \caption{Best performance on YCSB-A, when tuning SHAP's top-8 knobs; hand-picked top-8 knobs, and all knobs are baselines (left plot). Best performance on TPC-C, when tuning YCSB-A's top-8 knobs (SHAP, hand-picked) (right plot).\protect\footnotemark}
    \vspace{-1em}
\end{figure}

\footnotetext{Note that y-axis limits are chosen to improve readability of graphs. The only point below Y-axis minimum is the default configuration which is iteration 0.\label{note3}}

However, recent studies have shown that tuning a handful of knobs can be sufficient to achieve near-optimal performance and significantly reducing the number of knobs can accelerate the tuning process~\cite{kanellis2020too, van2021inquiry, zhang2021facilitating}.
Intuitively, the configuration search space generated by few (important) knobs is much smaller than the one when considering all knobs, making BO-based methods much more effective at finding configurations that yield good performance~\cite{shahriari2015taking, van2021inquiry}.
However, it still remains challenging to identify and select the \emph{correct} set of important knobs for different workloads or systems.

\subsubsection{Identifying Important Knobs}
All existing methods for automatically identifying important knobs use a ranking-based process~\cite{kanellis2020too, van2021inquiry, zhang2021facilitating}. Ranking-based methods typically use a space-filling sampling, like Latin Hypercube Sampling (LHS)~\cite{lhs}, to generate and evaluate thousands of configurations.
With this set of evaluated configurations alongside the measured performance, an ML model is trained to quantify how each knob affects the DBMS performance, an i.e., \textit{importance} score. 
Recent work~\cite{zhang2021facilitating} has shown that SHAP~\cite{shap} values provide the most meaningful importance score in the context of DBMS tuning.
SHAP uses a game-theoretic approach to break down the impact of each individual feature when analyzing the performance deviation from the default configuration.

Although these statistical methods do not require any human input for computing the important knob ranking, they require evaluating thousands of data samples~\cite{kanellis2020too, van2021inquiry}.
Thus, using important knob ranking to prune the configuration space is not useful in a setting where we start with zero knowledge.
Moreover, as we discuss next, methods that identify important knobs are \emph{(1)} not always reliable, which can lead to finding worse configurations, and \emph{(2)} the important knobs found for one workload may not work for others.

\vspace{1ex}
\noindent\textbf{Limitations of ranking important knobs.}
Using the read-write balanced YCSB-A~\cite{ycsb} as our workload, we generate (using LHS) and evaluate $2,500$ configurations for PostgreSQL v9.6, where we vary the values of $90$ knobs. Following prior work~\cite{zhang2021facilitating}, we employ SHAP (using RF) to rank all knobs in descending order of importance.
From this ranking, we select the top \textit{eight} ones, which, according to SHAP, should be able to provide most of the tuning performance gains.
We also consider two baselines: the full set of $90$ knobs, and a hand-picked set of \textit{eight} important knobs, which differs slightly from the SHAP-based one.
Table~\ref{tab:top_8_shap_vs_manual} shows both important knob sets.

Given these three sets, we launch a tuning session using SMAC for $100$ iterations, optimizing for throughput; we repeat the experiment five times with different random seeds.
\autoref{fig:top_8_smac} shows the best throughput reached at each iteration; shaded regions correspond to [5\%, 95\%] confidence intervals.
We make two observations.
First, tuning SHAP's top-8 knobs achieves worse final throughput compared to both the hand-picked top-8 knobs, and all knobs.
Thus, while it is possible to reduce the number of knobs tuned and still achieve optimal performance (e.g., hand-picked top-8 knobs), SHAP's ranking fails to extract the ``correct'' set of important knobs.
Second, tuning the hand-picked top-8 set of knobs identifies a better-performing configuration, while also converging much \textit{faster}.
Therefore, exploring the appropriate low-dimensional configuration space, can greatly benefit the performance of BO methods.

We next study the issue of transferring important knobs across workloads.
Using SHAP's important knobs for YCSB-A, we launch a similar experiment for a different workload, TPC-C
(Figure~\ref{fig:tpcc_top_8_ycsbA}).
\revision{We observe that when tuning all knobs, SMAC reaches a significantly higher throughout compared to only tuning the top-8 knobs from YCSB-A.}
Thus, reusing a set of important knobs derived from one workload to tune another, does not always lead to gains.

In summary, we see that tuning a low-dimensional space, generated by a set of important knobs can lead to finding better configurations with fewer iterations compared to the state-of-the-art SMAC algorithm.
However, existing statistical-based methods for selecting important knobs cannot be utilized as they are expensive, not always reliable and can lead to worse tuning outcomes. 

\section{Low-Dimensional Tuning}

In this section, we present a new approach which bypasses the need to identify the exact set of important knobs, yet realizes the benefits of low-dimensional tuning.

\noindent\textbf{Setup.}
We consider a scenario where a user/operator is tuning a DBMS for a specific workload given no prior knowledge.
Given $D$ tuning knobs, a $D$-dimensional space $X_D$ is constructed to represent the configuration search space.
Each point $p \in X_D$ corresponds to a specific DBMS configuration, where $p$'s coordinates determine the value of each of the $D$ tuning knobs.
This space is then given as input to a BO optimizer (e.g., SMAC), whose goal is to model the DBMS performance at each point $p \in X_D$. 
Due to the high dimensionality $D$ of the input search space $X_D$, modeling the DBMS performance across the whole space accurately requires the evaluation of many points (configurations).
Yet, as we've seen, many of those parameters those dimensions represent are less important, and hence amenable to compression in a model.
With this in mind, if instead of the input space $X_D$, the BO method received as an input, a smaller $d$-dimensional space $X_d$, where $d << D$, then fewer points would be needed to effectively learn the (smaller) space $X_d$.
Thus, choosing a smaller space $X_d$ instead of $X_D$ has the potential to improve the optimizer performance: i.e., find \textit{better} configurations or similar-performing ones, \textit{faster}.

However, not every reduced space $X_d$ may lead to benefits.
For example, tuning an $1$-dimensional space generated by a single \textit{non}-important knob would not lead to any performance improvements.
Therefore, it is vital for this smaller space $X_d$ to include DBMS configurations that can achieve high performance; ideally the best-performing ones.
Given this, we set the following goal:

\textit{For a user-provided $D$-dimensional input configuration space $X_D$, devise an \textit{approximation} using some $d$-dimensional space $X_d$, where \mbox{$d << D$}, which is likely to contain at least one point $p' \in X_d$ that can yield performance close to the optimal $p^* \in X_D$.}

\subsection{Synthetic Search Spaces}

The low dimensional spaces we discussed in Section~\ref{subsec:important-knobs} assumed that each dimension of the space corresponds to a specific configuration knob.
While this makes it easier for us to envision this space, it makes no difference for the optimizer, which can be given any search space as input.
To this end, we now \textit{decouple} this one-to-one relationship of configuration knobs to input search space dimensions.
In particular, we can use ``artificial'' dimensions (or \textit{synthetic} knobs) to generate our low dimensional space, $X_d$.
These synthetic knobs do not have any physical meaning themselves, yet their values determine the values of one (or more) DBMS configuration knobs.
In other words, one can define a \textit{mapping} from the synthetic knob values (i.e., approximated space $X_d$), to the physical DBMS configuration knobs (i.e., original input space $X_D$).
This enables us to realize the benefits of optimizing a low-dimensional space, while avoiding the need to identify important knobs.

The idea of using a synthetic smaller space $X_d$ to implicitly optimize a high-dimensional input search space $X_D$, has been recently studied in the BO community~\cite{pmlr-v161-eriksson21a,rembo}.
Multiple theoretically-sound approaches have been proposed to define this space, construct the mapping between the two spaces $X_d$, $X_D$, and/or extend the standard BO algorithm to better navigate through this smaller space~\cite{pmlr-v161-eriksson21a, rembo, hesbo, Letham2020alebo}.
For these methods to be effective, the target objective function should be affected by a small subset of the original high-dimensional space. If $d_e$ is the size of this subset, for a low-dimensional space $X_d$ with $d > d_e$, these methods provide theoretical guarantees that with high probability, each $p \in X_D$ can be obtained by projecting some point $p' \in X_d$.
In other words, if the low-dimensional space is \textit{larger} than the effective dimensionality of the objective function, then it is possible to achieve near-optimal performance just by tuning this smaller space~\cite{rembo, Letham2020alebo}.
As we discussed in Section~\ref{subsec:important-knobs}, a small set of important knobs are responsible for most of the DBMS performance improvements, which satisfies the necessary condition.
Thus, we believe that low-dimensional BO-methods are a promising way to improve the optimizer performance for database tuning.
We next discuss two popular methods.

\subsection{Random Low-Dimensional Projections}
\label{sec:rembo-hesbo}
REMBO~\cite{rembo} was one of the first attempts to solve the problem of high-dimensional search spaces in BO using \emph{randomized} low-dimensional linear projections.
Given as an input the target dimension $d$, REMBO first defines a $d$-dimensional search space, $X_d = [-\sqrt{d}, \sqrt{d}]^d$.
This search space is given as input to the BO method, which will suggest points $p \in X_d$.
Then, REMBO generates a random projection matrix $\mathbf{A} \in \mathbb{R}^{D~\times~d}$, whose entries are i.i.d. using $N(0,1)$.
Matrix $\mathbf{A}$ is then used to project a point suggested by the BO, $p$, from the low-dimensional $X_d$, to a corresponding point $\mathbf{\hat{p}}$ of the original high-dimensional space $X_D$.
To perform this projection a single matrix-vector product is computed as
$\mathbf{\hat{p}} = \mathbf{A}\mathbf{p}$.
Essentially, each synthetic knob $j$ of the $X_d$ space controls every knob $i$ of the $X_D$ space to a certain degree, quantified by the weight $A_{i, j}$ of the projection matrix (i.e., all-to-all mapping).

If the original space $X_D$ is unbounded (i.e., has no constraints), then the projected point $\mathbf{\hat{p}}$ will be a valid point in $X_D$.
Yet, in many problems (including DBMS tuning), there exist box constraints associated with $X_D$ (e.g., ranges of DBMS configuration knob values).
In this case, it is possible that for some point $p \in X_d$, the projected point $\mathbf{A}\mathbf{p} \notin X_D$.
REMBO handles this by ``clipping'' the projected point $\hat{p}$ to the \textit{nearest} point that belongs in $X_D$.
If say $X_D = [-1, 1]^D$, REMBO proposes to clip each point coordinate so that it lies within the $[-1, 1]$ range~\cite{rembo}.
However, subsequent works have found that this heuristic forces the optimization to be done on the facets of $X_D$, thus neglecting almost all interior points~\cite{Letham2020alebo}.

\begin{algorithm}[t]
\caption{BO with low-dimensional random projections. Blue \mbox{underlined} text highlights differences to original BO algorithm.}
\label{alg:modified_bo}
\renewcommand{\algorithmicrequire}{\textbf{Input:}}
\renewcommand{\algorithmicensure}{\textbf{Output:}}
\newcommand{\algodiff}[1]{{\color{blue}{\underline{#1}}}}
\begin{algorithmic}[1]
\Require $d, D, n_{init}, N_{iters}$
\Ensure Approximate optimizer $p*$

\State \algodiff{Generate random projection matrix $\mathbf{A} \in \mathbb{R}^{D~\times~d}$}
\State Generate $n_{init}$ points $\mathbf{p}_i \in X_d$ using a space-filling design (LHS)
\State Evaluate obj. function $f(\algodiff{\mathbf{A}\mathbf{p}_i)}$ for the generated points 
\State Fit initial data $\mathcal{D}_0 = \{(\mathbf{p}_i, f(\algodiff{\mathbf{A}\mathbf{p}_i)}\}_{i=1}^{n_{init}}$ to BO surrogate model
\For{$j = 1, ..., N_{iters}$}
    \State Find point \algodiff{$p_j \in X_d$} that maximizes the acquisition function
    \State Evaluate obj. function on the \algodiff{projected point $f(\mathbf{A}\mathbf{p}_j)$}
    \State Update surrogate model with $\mathcal{D}_j = \mathcal{D}_{j-1} \cup \{(\mathbf{p}_j, f(\algodiff{\mathbf{A}\mathbf{p}_j})\}$
\EndFor
\State \Return \algodiff{$\mathbf{A}\mathbf{p^*}$} for the best point $\mathbf{p}^* \in \mathcal{D}$
\end{algorithmic}
\end{algorithm}

\begin{figure}[t]
    \begin{minipage}{0.48\linewidth}
        \centering
        \includegraphics[width=\linewidth]{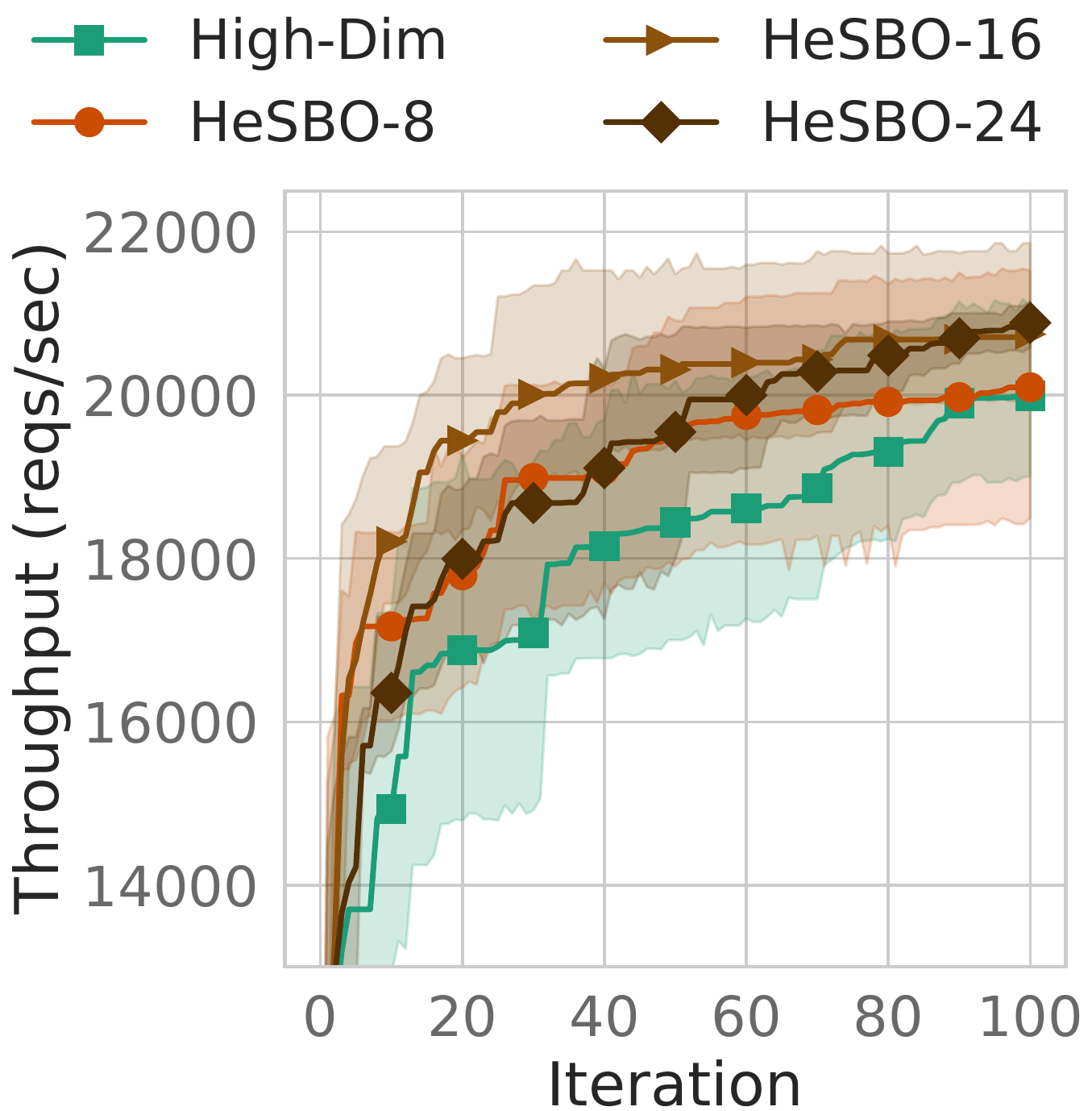}
        \vspace{-4ex}
        \captionsetup{labelformat=empty}
        \caption{HeSBO ($d=8, 16, 24$)}
    \end{minipage}
    \begin{minipage}{0.48\linewidth}
        \centering
        \includegraphics[width=\linewidth]{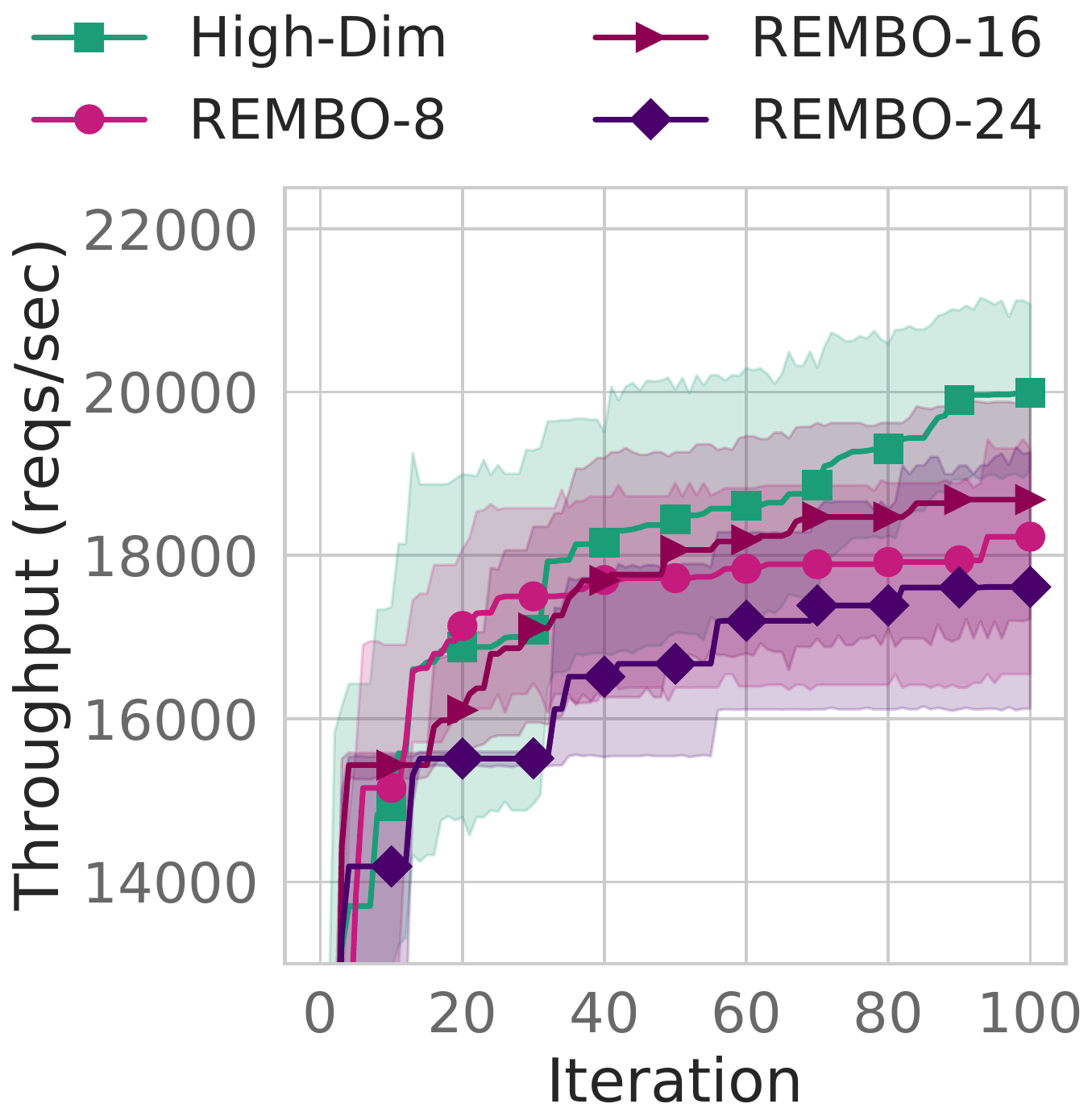}
        \vspace{-4ex}
        \captionsetup{labelformat=empty}
        \caption{REMBO ($d=8, 16, 24$)}
    \end{minipage}
    \addtocounter{figure}{-2}
    \vspace{-2ex}
    \caption{\revisionenv Best throughput on YCSB-A, when using SMAC to optimize a low-dimensional space from REMBO or HeSBO projections. Baseline (green line) tunes the original space. Shaded areas correspond to $95\%$ confidence interval ranges.\textsuperscript{\ref{note3}}}.
    \label{fig:lde_vs_orig_space}
    \vspace{-3ex}
\end{figure}

\vspace{-2ex}
\noindent\textbf{HeSBO.} To alleviate this drawback, researchers proposed HeSBO (i.e., Hashing-enhanced Subspace BO)~\cite{hesbo}, a random linear projection variant that avoids clipping.
\revision{HeSBO draws inspiration from hashing and sketching to construct its mapping.}
Assuming an original $D$-dimensional search space $X_D$ and given as an input the target dimension $d$, HeSBO first defines a $d$-dimensional search space $X_d = [-1, 1]^d$.
Then, it generates a random projection matrix $\mathbf{A} \in \mathbb{R}^{D~\times~d}$, such that each row contains exactly one non-zero element in a random column that is set to $\pm1$; the index of the column and the sign of the value are sampled uniformly at random.
\revision{It does so, using two uniform hashing functions: $h$ that chooses the non-zero entry for each dimension of $X_D$, and $\sigma$ that determines the sign of this entry.}
\revision{Intuitively, $\mathbf{A}$ defines an instance of count-sketch projection~\cite{charikar2002finding}, which is able to adequately preserve the characteristics (i.e., pairwise point distances) of the original high-dimensional space for up to $d$ (important) dimensions~\cite{hesbo}.}

HeSBO, as expressed by $\mathbf{A}$, provides an one-to-many mapping.
Every original knob $i$ in $X_D$ is controlled by \textit{exactly} one synthetic knob $j$ in $X_d$, while each synthetic knob can control multiple original ones.
Thus, it is impossible for any projection to fall outside of $X_D$.
\revision{This enables HeSBO to perform better than REMBO, especially when high-performing points are in the interior of $X_D$~\cite{Letham2020alebo, hesbo}.}

\subsection{BO with Random Projections}
\label{sec:3-bo-with-random-projections}

Integrating either REMBO or HeSBO in the original BO algorithm requires minimal changes.
Algorithm~\ref{alg:modified_bo} describes the modified version and highlights the main differences compared to the ``vanilla'' high-dimensional BO algorithm.
The user provides $d$, the number of dimensions of the low dimensional space.
Ideally, this number should be an estimate of the number of important knobs of the DBMS (we discuss how to choose this later).
The user also provides $n_{init}$, the number of initial random samples to be generated, as well as $N_{iters}$, the number of iterations to perform (time budget).
Reasonable defaults can also be provided based on prior studies.

At first, the projection matrix $\mathbf{A}$ is generated randomly, as described in the previous paragraphs.
Note that this matrix $\mathbf{A}$ is computed only once, and remains constant for the entire duration of the optimization process.
Then, given $n_{init}$, a space-filling sampling method (e.g., LHS) is employed that generates an initial set of points $p \in X_d$.
Since these points belong to the approximated space $X_d$, they are first projected to $X_D$ and then clipped (if using REMBO).
For each projected (and possibly clipped) point, the corresponding objective function is computed (i.e., workload is run).
Once all initial points have been evaluated, the points and the respective objective function values are used to initialize the surrogate model.

Now, the BO surrogate model can guide the optimization process.
Initially, the model suggests a candidate point $p \in X_d$ that maximizes the value of the acquisition function.
Recall from earlier that the acquisition function tries to propose points that either can provide knowledge about unknown regions (i.e., exploration), or improve the knowledge of already-found good regions, by exploring nearby regions (i.e., exploitation).
The suggested point is then projected (and possibly clipped) to $X_D$, as before.
Finally, the objective function is evaluated at that point, and the surrogate model parameters are updated using the new observation.
The above process is repeated, until we exhaust the given number of iterations.

\vspace{1ex}
\noindent\textbf{Converting Points to DBMS Knob Values.}
\label{sec:low-to-high-conversion}
So far, we assumed that both REMBO and HeSBO project the point suggested by the BO method to the uniform high-dimensional space $X_D = [-1, 1]^D$.
However, in reality a DBMS configuration knob space is much more heterogeneous compared to $X_D$.
Usually, a DBMS configuration space consists of many knobs of different types, like numerical or categorical.
The former receive values from a $[min, max]$ range of valid values, while the latter from a predefined list of choices (e.g., on, off).
Because the $[min, max]$ range of values (or the number of choices) can significantly vary across different knobs~\cite{ottertune}, we need to make sure that $[-1, 1]$ values are properly converted to meaningful values for the DBMS knobs.

To achieve this, we employ min-max uniform scaling.
In general, to translate a value $x \in [x_{min}, x_{max}]$ to a value $y \in [y_{min}, y_{max}]$, the following formula is used:
$y = y_{min} + \frac{x - x_{min}}{x_{max} - x_{min}} (y_{max} - y_{min})$.
Hence, for numerical knobs, we uniformly scale the projected value $x \in [-1, 1]$ to the respective $[min, max]$ range, of each knob.
If a numerical knob takes discrete values, we further round the scaled value to the correct integer value.
For categorical knobs,we perform a two-step conversion process.
We first rescale the projected value $x \in [-1, 1]$ to a value $y \in [0, 1]$, and then we split this range equally to as many bins, as is the number of different choices.
The final categorical knob value is defined by which bin the value $y$ falls into.

\subsection{Dimensionality of the Projected Space}
\label{sec:dim-and-case-study}

The only additional input for using these low dimensional projection methods is the number dimensions $d$.
Given that the objective function is unknown, it is usually not trivial to provide a good value for $d$.
Fortunately, the DBMS knob configuration tuning problem has been studied extensively, and we can rely on observations of past works to provide a good estimate for $d$~\cite{van2021inquiry, zhang2021facilitating, kanellis2020too}.

Prior works have shown good performance improvements when tuning around the $10$-$20\%$ most important, of all knobs considered for tuning.
For instance, in~\cite{kanellis2020too} the authors consider a subset of $30$ knobs for PostgreSQL v9.6, and show that tuning the $5$ most important ones ($\approx17\%$) can yield near-optimal performance.
Similarly, prior work~\cite{van2021inquiry} in tuning Oracle v12.2 used
a set of 10-40 hand-chosen knobs and~\cite{zhang2021facilitating} showed that
tuning the top-$20$, out of $197$ knobs ($\approx10\%$) was sufficient for MySQL v5.7.
Since the total number of knobs we consider for PostgreSQL v9.6 is $90$, we consider $d$ to be in the range of $[8, 16]$.
However, as we show next, setting $d$ to an even higher value (i.e., $>16$) is not detrimental to the optimizer performance.

\noindent\textbf{Case Study.}
We perform a case study using YCSB-A.
Our goal is two-fold.
First, we want to evaluate how effective REMBO and HeSBO methods are at finding a good projection of the original DBMS configuration space, which can result in better performance.
Second, we want to explore how much our selection of $d$ influences the optimizer performance.
Our setup consists of tuning $90$ knobs of PostgreSQL v9.6 for $100$ iterations (first $10$ are LHS-generated samples) targeting maximum throughput using SMAC (more setup details in Section~\ref{sec:eval}).
We experiment with both REMBO and HeSBO, where we set $d$ to $8$, $16$, or $24$ dimensions.

~\autoref{fig:lde_vs_orig_space} shows the best throughput achieved at each iteration.
Overall, we observe that HeSBO manages to outperform the baseline for all values of $d$, while REMBO is not able to match that performance.
REMBO finishes the tuning session with a final configuration that achieves $10$-$15\%$ worse than the baseline.
We find that this is because most of the low-dimensional points are clipped during projection to the original DBMS configuration space, which makes it impossible for the optimizer to explore the interior points.

On the other hand, HeSBO  does not suffer from the aforementioned issue and performs much better for all values of $d$.
This shows that HeSBO can effectively project the original space to a smaller one that is easier to explore, while also retaining many points that achieve good performance.
Moreover, we see that $d$ does not greatly affect the quality of the projected space.
Even though the $16$-dimensional HeSBO performs the best, all three manage to end up with similar- or better- performing configurations (within $5\%$) at the end of the tuning session, while also converging much faster.
This highlights the robustness of HeSBO, and means that we can provide a reasonable estimate of the number of important knobs, rather than an exact number.
For the remainder of the paper, we use a $16$-dimensional HeSBO projection for our experiments.

\vspace{-2ex}
\section{Not All Knob Values Are Equal}

Shrinking the search space by reducing the number of dimensions stemmed from our prior knowledge of a few important knobs affecting the DBMS performance. We next discuss how auto tuning frameworks can take advantage of another DBMS specific observation: not all knob values have the same performance semantics.

\subsection{Special Knob Values in Databases}
\label{subsec:special-knobs}

\begin{table*}[!htp]
\small
\centering
\caption{Three examples of hybrid knobs found in PostgreSQL; Each special value performs a different action each time.}
\label{tab:hybrid_knobs}
\vspace{-1em}
\begin{tabular}{llp{5.5cm}p{6.5cm}r}
\toprule
\textbf{Knob name} &\textbf{Range} &\textbf{Short Description} &\textbf{Special Value Action} \\
\midrule
\code{backend\_flush\_after} & $[0, 256]$ & \emph{Number of pages after which previously \mbox{performed} writes are flushed to disk.} & If set to 0, the forced writeback is \textbf{disabled}. \\
\code{geqo\_pool\_size} & $[0, 2^{32})$ & \emph{Controls the pool size used by GEQO, that is the number of individuals in the genetic population.} & If set to 0, then a \textbf{suitable value} is chosen based on \code{geqo\_effort} and the number of tables in the query. \\
\texttt{wal\_buffers} & $[-1, 2^{18})$ & \emph{Sets the number of disk-page buffers in shared memory for WAL} & If set to -1, a size equal to $\mathbf{1/32}$nd of \texttt{shared\_buffers} is \textbf{selected} ($\ge64$kB, not more than one WAL segment). \\
\bottomrule
\end{tabular}
\vspace{-0.1in}
\end{table*}

The configuration search space of a DBMS is heterogeneous, and consists of numerical and categorical knobs.
Categorical knobs are inherently different from numerical ones: they enable (or disable) some internal functionality of a component, or select mutually exclusive algorithms for a task.
For example, PostgreSQL exposes a plethora of such knobs (e.g., \texttt{enable\_sort}, \texttt{enable\_nestloop} etc.)~\cite{postgresql-documentation}.
Choosing different values for a categorical knob may result in substantial DBMS performance variations.
Hence, most efficient BO methods treat each categorical knob value independently and make no assumptions regarding their order~\cite{zhang2021facilitating, hutter2011smac}. 

On the other hand, the values of numerical knobs have a natural order, which is actively exploited by the BO methods when performing local search.
In particular, when the optimizer has identified a good-performing configuration (point), it often opts to explore the nearby regions in order to improve the current optimum~\cite{shahriari2015taking}.
However we find, in practice, some numerical knobs \textit{do} have special values (e.g., $-1$, $0$) that inevitably break this natural order~\cite{pavlo2021electricsheep}.
We term these knobs as \textit{hybrid} knobs.
If such a knob is set to its special value, it does something very different compared to what it normally does (e.g., disables some feature).
Otherwise, it behaves as a regular numerical knob where it sets some underlying functionality of the respective component (e.g., buffer size, flush delays, etc.).

We identified $17$ hybrid knobs for PostgreSQL v9.6 ($24$ exist for PostgreSQL v13.6) from the official documentation~\cite{postgresql-documentation}; special values are explicitly mentioned in the knob description.
Interestingly enough, for about half of the hybrid knobs, the special value is used in the default configuration.
~\autoref{tab:hybrid_knobs} shows some interesting examples.
In general, most common behaviors related to special values include \emph{(i)} disabling some feature, \emph{(ii)} inferring this knob's value based on some other's knob, or \emph{(iii)} setting this knob's value using an internal heuristic (or to a predefined value).
It is worth noting that the existence of hybrid knobs is neither limited to this specific PostgreSQL version, nor to this DBMS alone.
Other popular systems like MySQL~\cite{mysql-documentation}, Apache Cassandra~\cite{apachecassandra-documentation} and even the Linux kernel~\cite{linux-kernel-docs, linux-kernel-docs-2} also expose similar knobs in their configuration.

\begin{figure}[tp]
    \begin{minipage}{0.40\linewidth}
        \includegraphics[width=\linewidth, trim={0 0.15in 0 0 }, clip]{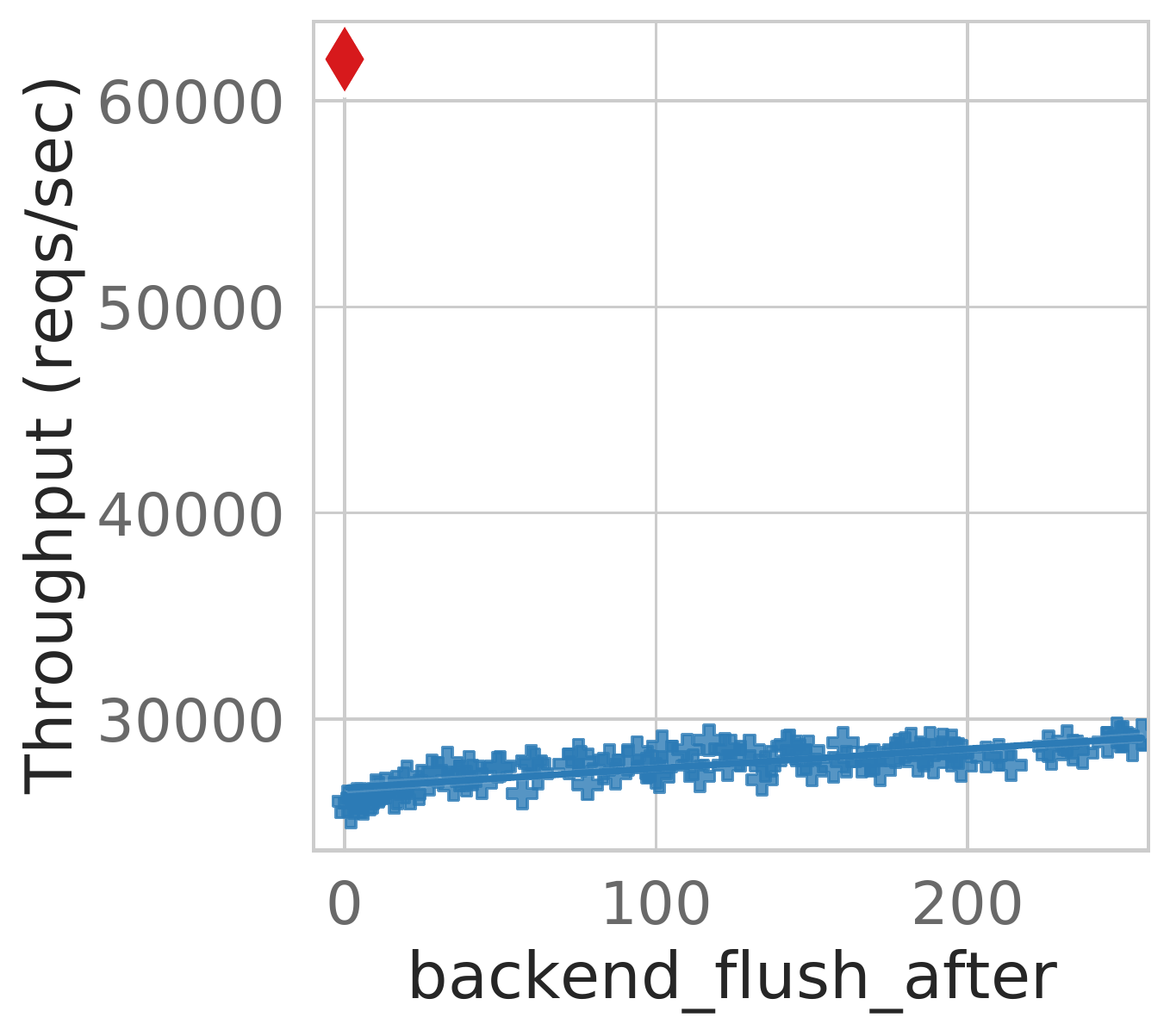}
        \vspace{-0.25in}
        \captionsetup{singlelinecheck=false,justification=raggedright}
        \caption{Effect on perf. of \mbox{special} value ``$0$''.}
        \label{fig:backend_flush_after_scan}
    \end{minipage}
    \hspace{0.02in}
    \begin{minipage}{0.57\linewidth}
    \begin{minipage}{0.34\linewidth}
        \footnotesize
        \begin{tabular}{l}
         BO Suggests \\ Value
         $\in$\texttt{[0,MAX]}\\
         \\
         Scale to \texttt{[0,1]} \\ \\
         Apply Special \\ Value Bias $p$ \\
         \\
         Final Value 
        \end{tabular}
    \end{minipage}
    \hspace{0.01in}
    \begin{minipage}{0.63\linewidth}
        \includegraphics[width=\linewidth,trim={0 0 0 1em}, clip]{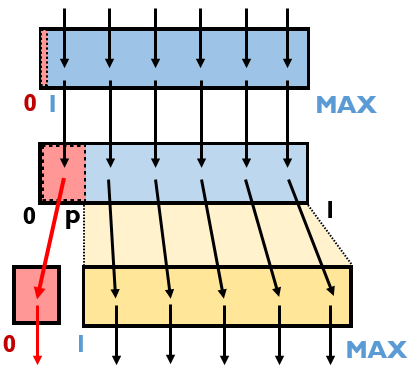}
    \end{minipage}
    \caption{Applying bias to special value ``$0$'' of a hybrid knob.}
    \label{fig:bias_sampling}
    \end{minipage}
    \vspace{-3ex}
\end{figure}

Special values make the modeling of the DBMS performance more difficult because they represent a discontinuity in objective function output relative to its input.
Figure~\ref{fig:backend_flush_after_scan} shows such an example for PostgreSQL's \texttt{backend\_flush\_after} knob, when executing a read-heavy YCSB workload: YCSB-B.
Typically, larger values for this knob translates to allowing more bytes to remain in kernel page cache, before issuing a writeback request.
A value of ``$0$'' (red diamond in the figure), however, disables the writeback mechanism altogether, resulting in far higher throughput for this workload.
From the performance numbers in this figure, we can also deduce the following: if the optimizer tuning this knob lacked the knowledge about the special meaning of value $0$, it would be unlikely that it would ever choose this value during tuning.
That is because $0$ is located near the worse-performing values (i.e., 1-10).
Instead, the optimizer would focus on exploring the ``neighborhood'' of larger values for this knob, which seem more promising.
Another example is when the number of possible values for a hybrid knob is very large; in this case, it is unlikely that the special value will be among the initial set of random values used to bootstrap the optimizer.
For example, assuming an initial set of $10$ uniformly sampled points, the probability of the special value being chosen at least once for the aforementioned \texttt{backend\_flush\_after} knob is less than $4\%$ \footnote{$10$ independent Bernoulli trials, each with \mbox{$1/(256 + 1)$} chance of success}.

\noindent
\textbf{Biasing Special Value Sampling Probability}
We propose a methodology that enables the optimizer to observe the effect of a special value early on in the tuning process.
Our key idea is to change the hybrid knobs values proposed, such that we \textit{bias} the probability of the special value being evaluated.
To do this we associate a fixed probability ($p$) for sampling a special value before we begin tuning.
To simplify our design, we keep the same probability $p$ for all hybrid knobs we tune.
Figure~\ref{fig:bias_sampling} illustrates this process for a single hybrid knob that takes values in $[0, max]$ range, where ``$0$'' is the special value.
Initially, a value $v \in [0, max]$ for this knob is suggested by the optimizer, or through random search.
Given this value $v$, we then uniformly scale it to the $[0, 1]$ range (using the scaling method described in Section~\ref{sec:low-to-high-conversion}).
If this scaled value is the range $[0,p)$, then we set the resulting knob value $v$ to the special value (i.e., $v$= 0), instead.
Otherwise, we uniformly scale it back to the original, minus the special value, range (i.e., $[1,max]$).
Finally, the resulting value is used at the DBMS configuration to be evaluated.
Note that this methodology requires no modifications to the underlying optimizer, as it only takes place \textit{after} a certain suggestion is made; thus it can be combined with any optimizer.
Furthermore, an extension for hybrid knobs with multiple special values is straightforward: we can define multiple $p_i$ (possibly equal with each other), where each one biases exactly one special value.

\noindent{\textbf{Choosing the Amount of Bias.}}
Evaluating the impact of the special value early on in the tuning session can help improve total tuning time.
Any existent performance anomalies caused by special values can be discovered and the optimizer can then leverage this knowledge. 
While it might be tempting to construct a configuration consisting solely of special values, and evaluate its performance immediately, this would not help, as all potential anomalies may happen simultaneously.
Thus, even if there exists a significant performance improvement from one special value, the optimizer might not be able to attribute this gain.
Alternatively, allocating an iteration to measure the effect of every special value independently, may be sub-optimal, especially if the number of such knobs is large.

Suppose the initial set of random configurations consists of $n_{init}$ samples and we set the amount of bias as $p$, the number of samples that evaluate the special value follows a binomial distribution $B(n_{init}, p)$.
By default, we choose this probability as 20\% as it leads to a $\sim$90\% confidence level of evaluating the special value (at least once) in the initial set of random configurations.
A nice property of this method is that it is invariant to the number of hybrid knobs that are tuned.
Users can also alter the value based on their preference.

Two things are worth noting before finishing our discussion regarding hybrid knobs.
First, even with a reasonably high confidence level, it is possible that a special value for a knob may not be evaluated during the initial iterations.
However, random configurations that are typically proposed by the optimizer periodically (e.g., every $n$-th iterations) are also biased towards each hybrid knobs' special value and this also handles any undesirable interactions across special values of different knobs.
Second, the special value bias does not obstruct the optimizer in any way when it performs local search.
Regardless of whether a special value improves performance or not, the optimizer is still able to fine-tune already-evaluated configurations to further improve performance.

\begin{figure}[tp]
    \begin{minipage}{0.47\linewidth}
        \includegraphics[width=\linewidth]{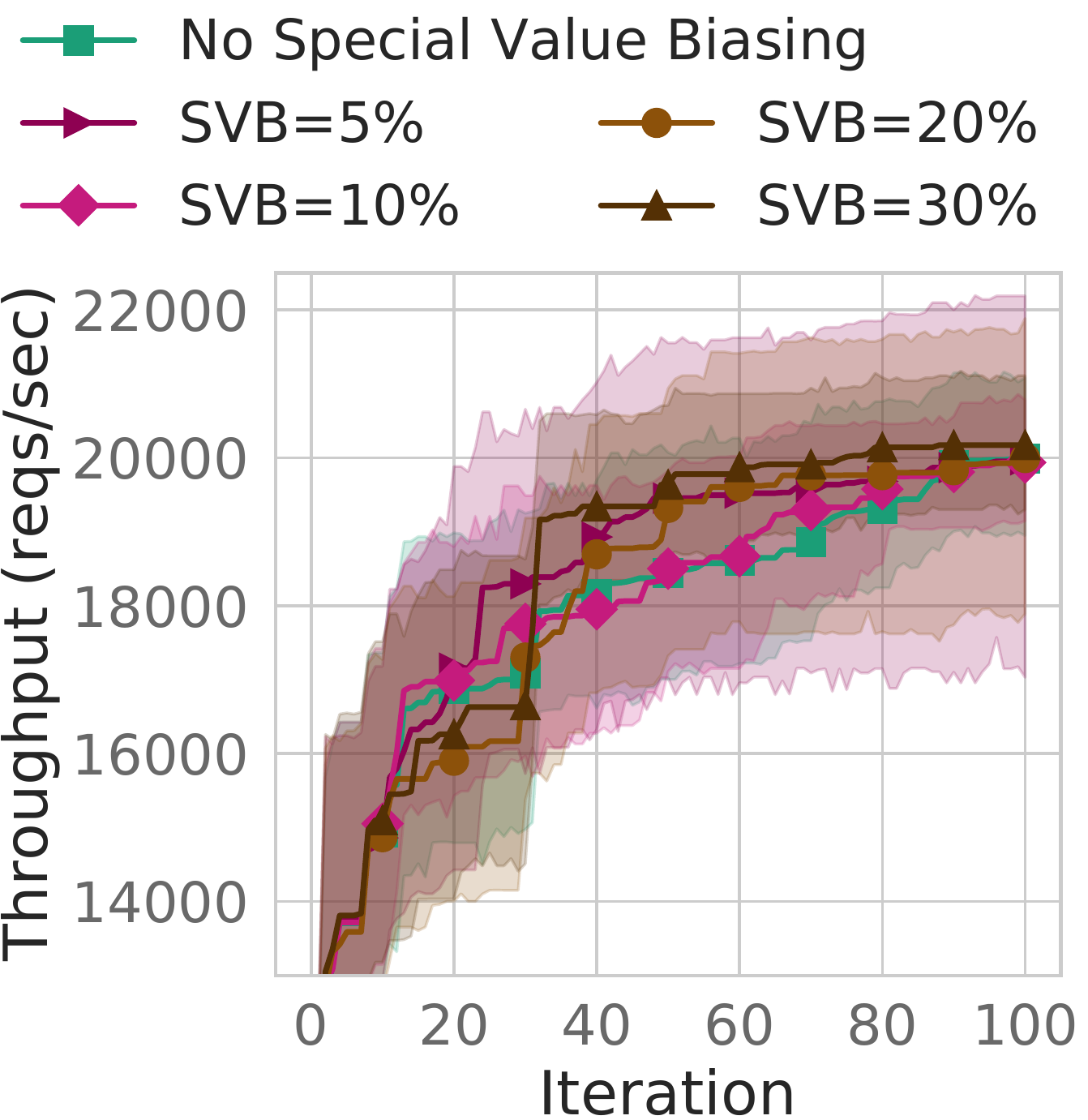}
    \end{minipage}
    \begin{minipage}{0.49\linewidth}
        \includegraphics[width=\linewidth]{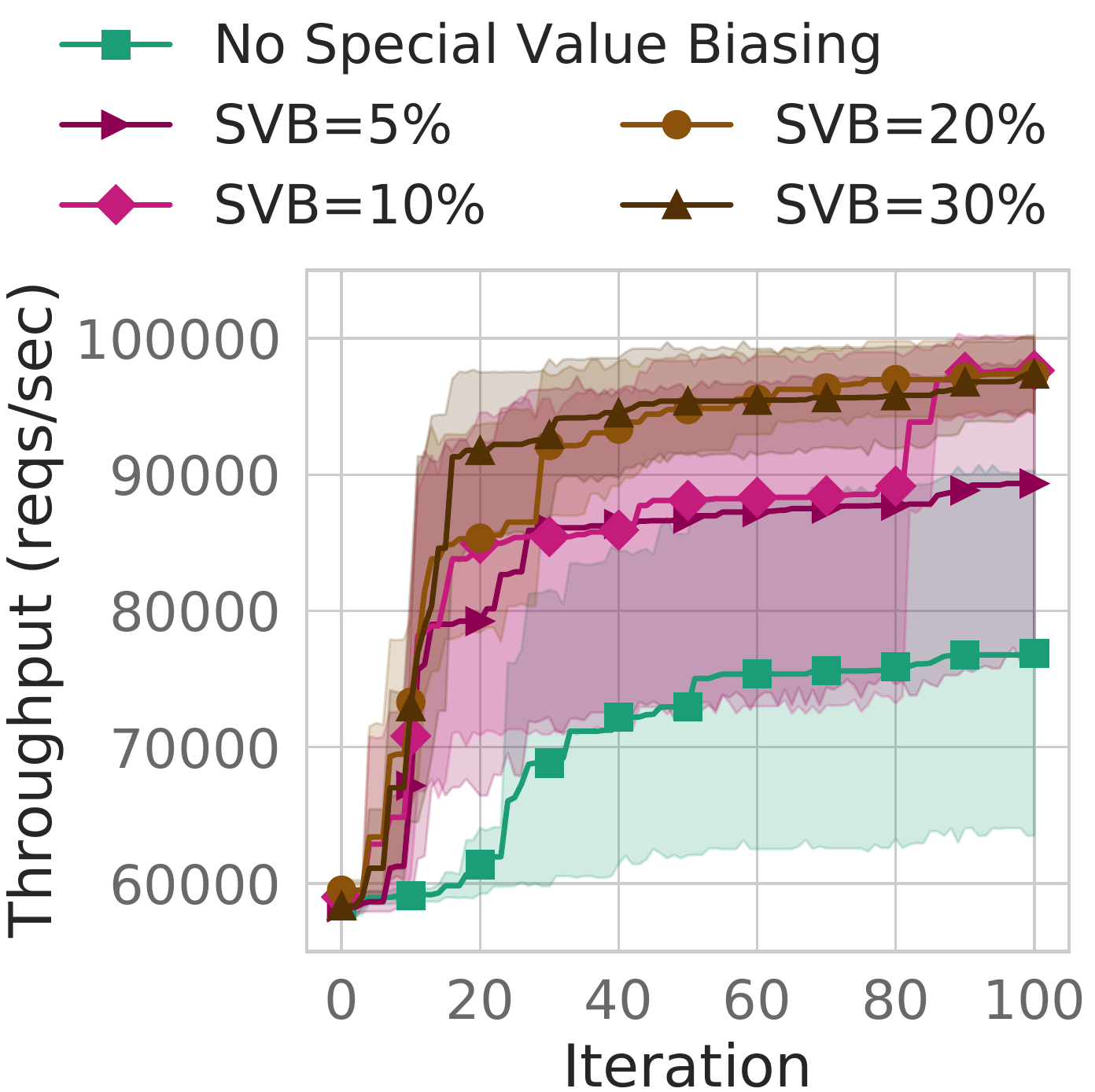}
    \end{minipage}
    \vspace{-2ex}
    \caption{\revision{Best performance achieved for YCSB-A (left plot) and YCSB-B (right plot) when applying special value biasing to hybrid knobs (top-left is better).\textsuperscript{\ref{note3}}}}
    \label{fig:bias_sampling_sensitivity}
    \vspace{-2em}
\end{figure}

{
\revisionenv
\noindent
\textbf{Case Study.}
We conduct an experiment with two workloads, YCSB-A, YCSB-B, in order to explore the impact of special value biasing to the performance of the optimizer.
We use the same setup used in the case study of Section~\ref{sec:dim-and-case-study} (PostgreSQL v9.6, SMAC, $10$ LHS-generated initial samples), and we vary the bias percentage from $5\%$ to $30\%$.
Figure~\ref{fig:bias_sampling_sensitivity} shows the best throughput reached at each iteration.
We observe that while for YCSB-A special value biasing does not provide any measurable gains, for YCSB-B the gains are large and increase with the amount of biasing applied up to $20\%$.
For the rest of the paper we use a $20\%$ bias percentage.
}

\begin{figure}[tp]
    \begin{minipage}{0.47\linewidth}
        \includegraphics[width=\linewidth]{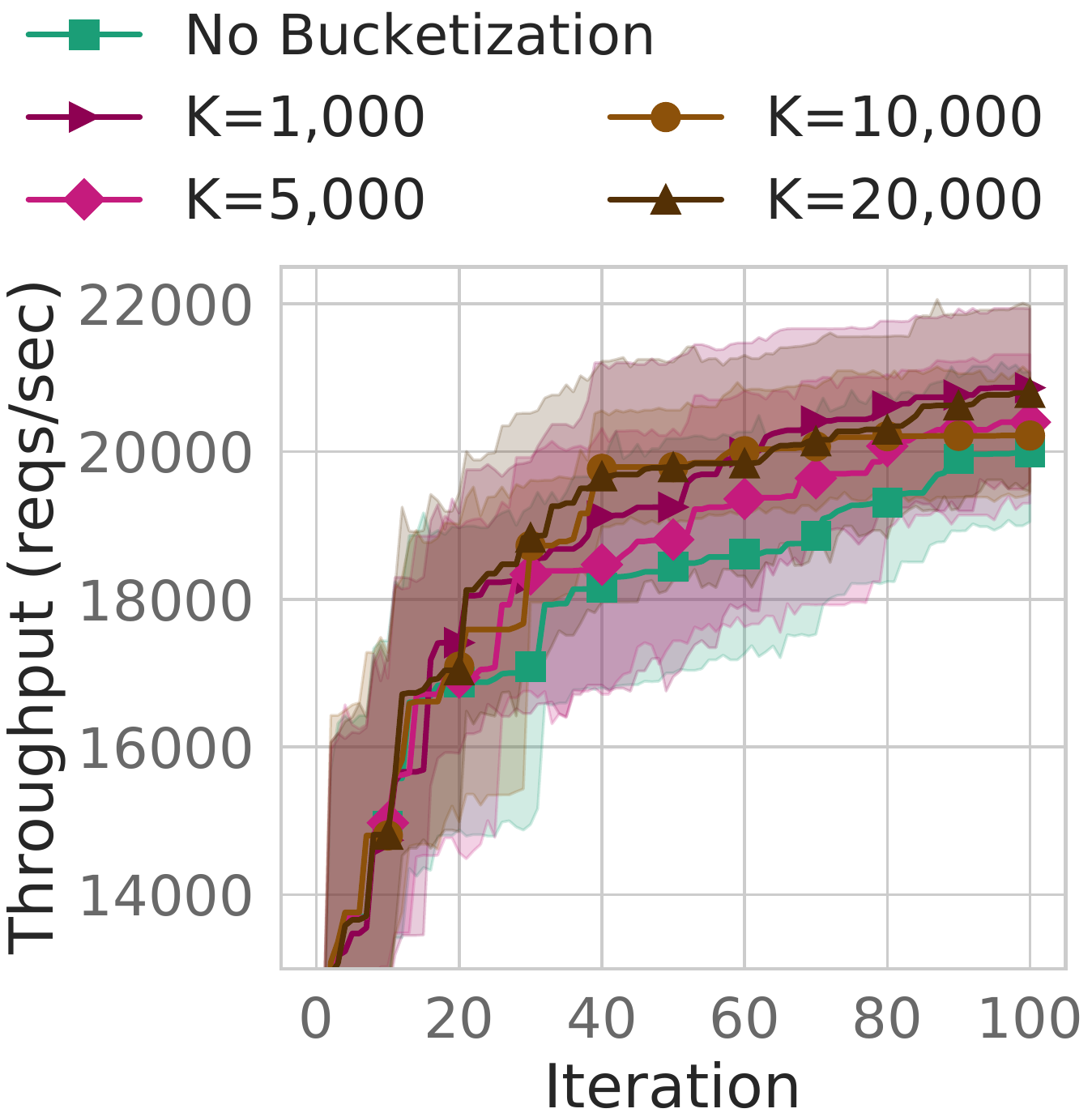}
    \end{minipage}
    \begin{minipage}{0.49\linewidth}
        \includegraphics[width=\linewidth]{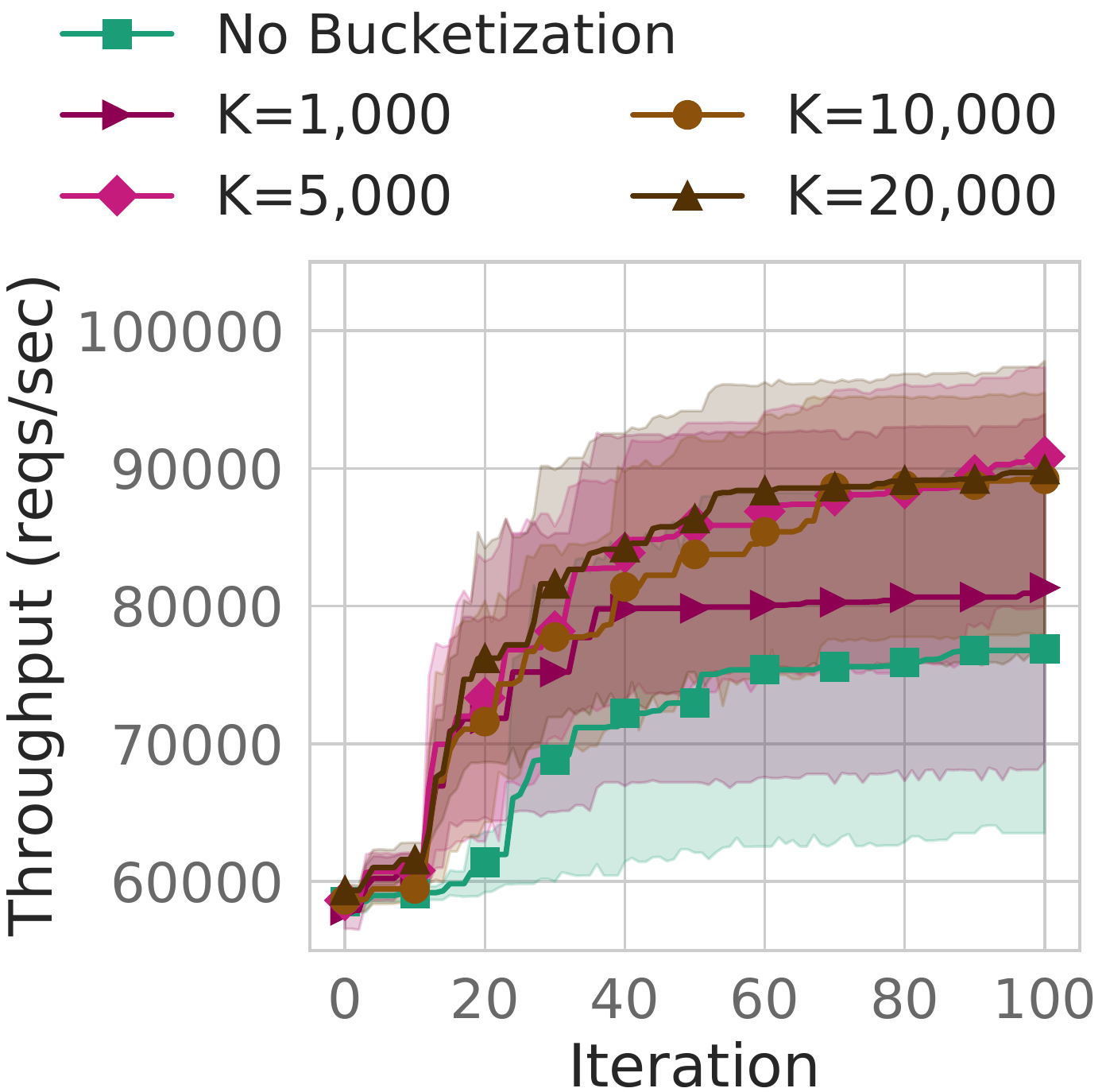}
    \end{minipage}
    \vspace{-2.5ex}
    \caption{\revision{Best performance achieved for YCSB-A (left plot) and YCSB-B (right plot) when tuning the \mbox{bucketized} space vs. the original space (top-left is better).\textsuperscript{\ref{note3}}}}
    \label{fig:bucketization-sensitivity}
    \vspace{-1em}
\end{figure}

\subsection{Configuration Space Bucketization}
\label{subsec:bucketing}

\begin{figure*}
    \centering
    \begin{minipage}{0.68\textwidth}
        \centering
        \small
        \captionof{table}{Examples of discrete knobs with large value range for PostgreSQL v9.6}
        \label{tab:bucketization}
        \vspace{-1em}
        \begin{tabular}{lllp{5cm}}\toprule
        \textbf{\small Knob name} & \textbf{\small Value Range} &\textbf{\small Step} &\textbf{\small Short description} \\
        \midrule
        commit\_delay &[0, 100000] &$1\mu$s & {\small Sets the delay in  microseconds between transaction commit and flushing WAL to disk} \\
        max\_files\_per\_process &[25, 50000$^*$] &1 & {\small Sets the maximum number of simultaneously open files for each server process} \\
        shared\_buffers &[128kB, 16GB$^*$] &8kB & {\small Sets the amount of memory the database server uses for shared memory buffers} \\
        wal\_writer\_flush\_after &[8kB, 16GB$^*$] &8kB & {\small Amount of WAL written out by WAL writer that triggers a flush} \\
        \bottomrule
        \end{tabular}
        \vspace{1pt}
        \footnotesize{\\\small{$^*$}The upper bound value for this knob is infinite, but we pruned it down to this more reasonable value.}
        \vspace{-0.1in}
    \end{minipage}
    \begin{minipage}{0.31\textwidth}
        \centering
        \captionof{table}{Workload Properties}
        \label{tab:workload}
        \vspace{-1em}
        \begin{tabular}{ccc}
        \toprule
        \multirow{2}{*}{\textbf{Workload}} & \multirow{2}{*}{\textbf{\begin{tabular}[c]{@{}c@{}}\# Tables \\ (\# Columns)\end{tabular}}} & \multirow{2}{*}{\textbf{\begin{tabular}[c]{@{}c@{}}RO \\ Txns\end{tabular}}} \\
         &  &  \\
        \midrule
        \textbf{YCSB-A} & 1 (11) & 50\% \\
        \textbf{YCSB-B} & 1 (11) & 95\% \\
        \textbf{TPC-C} & 9 (92) & 8\% \\
        \textbf{SEATS} & 10 (189) & 45\% \\
        \textbf{Twitter} & 5 (18) & 1\% \\
        \textbf{RS} & 4 (23) & 33\% \\
        \bottomrule
        \end{tabular}
    \end{minipage}
\end{figure*}

While biased sampling of special values can improve the exploration phase of the optimizer, we next look at how discrete configuration knobs in DBMS' can affect the exploitation (or local search) phase. 

Discrete configuration knobs exposed by the DBMS can have very dissimilar ranges.
Some knobs have $10$ possible values while others range in millions.
A broader range of values gives more fine-grained control of the underlying mechanism.
Table~\ref{tab:bucketization} lists some knobs in PostgreSQL that have a large number of unique values.
One example is \code{shared\_buffers}, which specifies the size of in-memory shared buffers at the granularity of a single page.
Another example is \code{commit\_delay}, which controls the delay before a committed transaction is flushed to disk; this delay can be set at microseconds granularity.
Again, the existence of knobs with such characteristics is not restricted to PostgreSQL~\cite{apachecassandra-documentation, mysql-documentation}.

While having many different choices for knob values enables fine-grained control, it greatly increases the size of the configuration search space.
We observe that for many configuration knobs with large values ranges, small changes are \emph{unlikely} to affect the DBMS performance significantly.
For instance, when tuning the value of \code{commit\_delay}, the performance is similar for $10,000\mu s$ and $10,100\mu s$ (i.e., adding $0.1ms$).
Based on this observation, we explore \emph{bucketizing} the knob value space at fixed intervals.
We can do this by limiting the number of unique values that any configuration knob can take to $K$.
Knobs with more than $K$ values are rounded down to use only $K$ unique values; these new values are uniformly spread across the entire range (i.e., if $K=100$ for \code{commit\_delay}, the new set of discrete values will be 0, 1000, 2000 ... 100000).
Knobs with less than $K$ values remain unaffected, and are tuned as before.

{
\revisionenv
\noindent\textbf{Case Study.}
We use a simple policy to set $K$ and explore the effect of bucketizing.
We chose the same $K$ for all knobs as it is hard to know the ideal value without performing expensive offline profiling.
We set $K$ based on the distribution of the ranges of values such that the number of knobs to be bucketized is $P\%$ of all knobs.
}

{
\revisionenv
We run an experiment with YCSB-A, YCSB-B workloads and compare SMAC's performance when tuning a bucketized space to tuning the original configuration space, for different values of $K$ (from $1,000$ to $20,000$).
Figures~\ref{fig:bucketization-sensitivity} shows the best throughput reached at each iteration.
For YCSB-A, while the performance in the first $20$ iterations is similar, in later iterations the bucketized (smaller) space reaches a better-performing configuration faster (for most values of $K$).
For YCSB-B we see larger benefits starting at iteration 10 (especially for $K \ge 5,000$).
Thus, we see that the benefits from bucketization varies across workloads.
In the future, we plan to better understand when BO methods benefit from using bucketization, especially given prior work in ML~\cite{JMLR:v13:bergstra12a} has shown the pitfalls of using a fixed grid for hyperparameter optimization.
For the rest of this paper we set $K=10,000$ (i.e., from $P\%=50\%$). 
}

\vspace{-1ex}
\section{LlamaTune Design}

So far, we have presented three DBMS specific observations that can improve the configuration optimizer: \emph{(i)} random low-dimensional projections, \emph{(ii)} biasing the special value of hybrid knobs, and \emph{(iii)} bucketization for knobs with many unique values.
We next present \sys{}, a \textit{unified} design that incorporates these approaches.

A unified design should satisfy three main requirements.
First, the optimizer should always operate on the low-dimensional space, and not in the original DBMS knob configuration one.
Second, the special value biasing should be performed \textit{only} over the set of hybrid knobs.
Otherwise, we might unnecessarily skew the values of other knobs towards non-existent special value(s).
Finally, the bucketized value space for the set of knobs should be exposed to the optimizer, so it is aware of the larger sampling intervals; otherwise it will still continue to sample at finer granularities.

In our attempt to meet the above design requirements, we devise a unified tuning pipeline (\sys{}) that employs a \textit{bucketized} version of the low-dimensional search space, and applies the special value biasing only after a point has been suggested.
Instead of the original low-dimensional continuous space $X_D = [-1, 1]^D$, we define a bucketized search space $X_D'$, similar to $X_D$, but with the ability to sample values only at fixed intervals, as defined by the number of maximum unique values (i.e., $K$).
Exposing the space $X_D'$ to the optimizer satisfies the first, but only partly the third requirement (we explain more below).
The second requirement is satisfied by applying the special value bias after the point has been projected to the original space, but before the values are re-scaled.
This way, the biasing approach does not interfere with the internals of the random projection method.
The main limitation of this design is that bucketizing the entire search space may affect fine tuning of continuous knobs.
However, we believe that the performance gains of optimizing a smaller space outweighs this drawback in most cases, especially when we use a conservative value for $K$.

\noindent\textbf{Example:}
In~\autoref{fig:tuner_design}, we present an example of our design.
We consider a scenario where we tune five DBMS knobs using a $2$-dimensional random projection derived by HeSBO.
We set the special value bias to $20\%$, and bucketize the low-dimensional space to $K=10,000$.
Among these knobs, \texttt{commit\_delay} is a discrete knob, \texttt{geqo\_selection\_bias} is continuous knob, and \texttt{enable\_seqscan} is a binary categorical knob.
The remaining two knobs are hybrid knobs: \texttt{backend\_flush\_after} has a special value of ``$0$''; for \texttt{wal\_buffers} it is ``$-1$''.

Initially, the optimizer suggests a point $[-0.8, 0.4]$ that belongs to the low-dim $[-1, 1]^d$ bucketized space.
The value $Q$ denotes the size of the fixed interval (i.e., $2/K$).
This point is then projected to a high-dimensional point in $[-1, 1]^D$ (i.e., $D=5$).
The first synthetic knob is mapped to both hybrid knobs (but with opposite $\pm$ signs), while the second one maps to the remaining three.
The projected point now corresponds to a real DBMS knob configuration.
Next, we normalize all knob values to $[0, 1]$, and apply special value biasing only for the two hybrid knob values.
Here, \texttt{backend\_flush\_after}'s original value of $0.1$ has been biased towards its special value $0$ (i.e., inside the $[0, 0.2)$ range), while \texttt{wal\_buffers} has been scaled (i.e., inside the $[0.2, 1]$ range) to compensate for the $20\%$ special value bias.
Finally, all normalized values are converted to the corresponding physical knob values.
These values now comprise the DBMS configuration that is evaluated on the real system.

\begin{figure*}[!tp]
    \centering
    \includegraphics[width=0.96\textwidth,trim={0 0 0 1em}, clip]{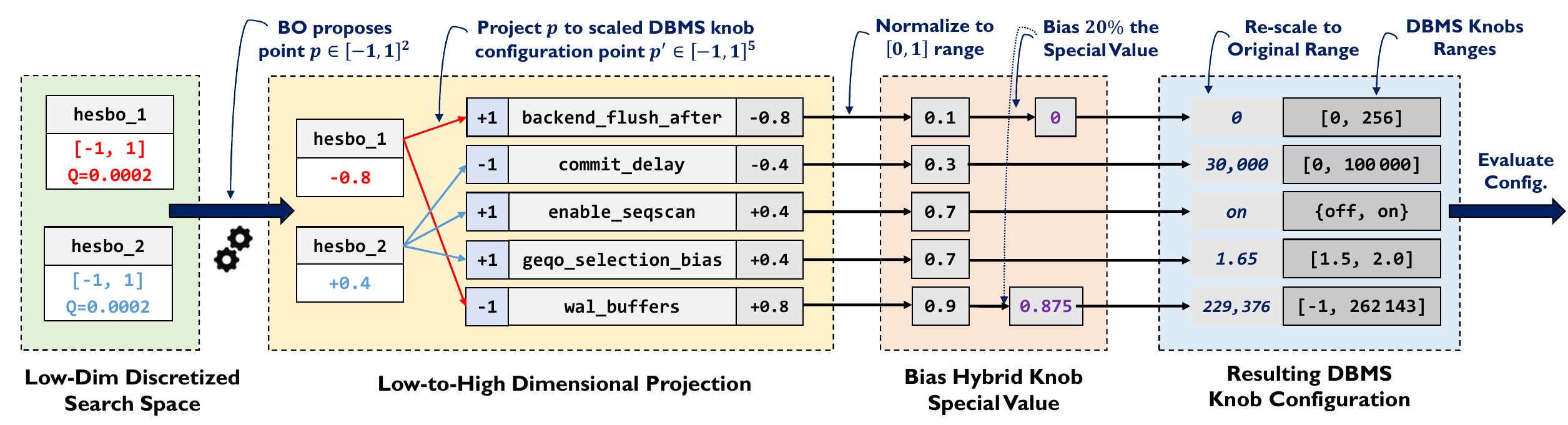}
    \vspace{-2ex}
    \caption{\sys{}: Tuning example that highlights the unified end-to-end pipeline.}
    \label{fig:tuner_design}
    \vspace{-1ex}
\end{figure*}


\vspace{-1ex}
\section{Evaluation}
\label{sec:eval}
In this section, we evaluate \sys on various workloads and show that it can improve the time taken for tuning PostgreSQL on six different workloads by $1.96\times$ up to $11.0\times$ when using SMAC.

\vspace{-1ex}
\subsection{Experiment Setup}

\noindent\textbf{Hardware.}
We conduct our experiments on CloudLab~\cite{cloudlab}.
Each tuning session runs on a single \texttt{c220g5} node located at the Wisconsin cluster.
The DBMS runs on top of a 10-core Intel Xeon Silver 4114 CPU with 16 GB of RAM and uses a 480 GB SATA SSD.
The workload clients and the optimizer algorithm are isolated from the DBMS, and run on a separate CPU (each node has two sockets).

\noindent\textbf{Tuning Settings.}
\revision{We run most experiments with PostgreSQL v9.6, but also use PostgreSQL v13.6}.
We select a set of tunable 90 knobs that could affect DBMS performance; we exclude from this set all knobs related to debugging, security, and path-setting.
We repeat each experiment \textit{five} times using different random seeds as input to our optimizer; we report the mean best performance from all runs.
Each tuning session consists of $100$ iterations
, and each run consists of five minutes of running the workload.
For most experiments we optimize for overall system throughput (i.e., higher is better); we also show that our approach can yield gains when optimizing for $95$-th percentile latency.
The configurations of the first $10$ iterations are generated randomly using LHS~\cite{lhs}.
For those configurations that cause the DBMS to crash, we assign one fourth of the worst throughput we have seen so far to that iteration; initially this value is set to the performance of the default configuration.
The above tuning configuration is similar to prior works~\cite{zhang2021facilitating, van2021inquiry, ottertune}.

\begin{table}[t]
\captionsetup{width=.49\textwidth}
\caption{Perf. gains of \sys{} when coupled with SMAC. We report average and $\mathbf{[5\%,95\%]}$ confidence interval numbers.}
\label{tab:throughput-smac}
\vspace{-1em}
\scalebox{0.885}{
\begin{tabular}{ccccc}
\toprule
\multirow{2}{*}{\textbf{Workload}} & \multicolumn{2}{c}{\textbf{\begin{tabular}[c]{@{}c@{}}Final Throughput \\ Improvement\end{tabular}}} & \multicolumn{2}{c}{\textbf{\begin{tabular}[c]{@{}c@{}}Time-to-Optimal\\ Speedup\end{tabular}}} \\
 & Average & {[}5\%, 95\%{]} CI & Average & {[}5\%, 95\%{]} CI \\
 \midrule
\textbf{YCSB-A} & 2.74\% & {[}0.9\%, 4.5\%{]} & 1.96x {[}51 iter{]} & {[}1.1x, 3.3x{]} \\
\textbf{YCSB-B} & 20.85\% & {[}10.2\%, 27.9\%{]} & 5.5x {[}18 iter{]} & {[}1.2x, 33.0x{]} \\
\textbf{TPC-C} & 6.30\% & {[}-0.3\%, 12.4\%{]} & 11.0x {[}9 iter{]} & {[}1.0x, 14.1x{]} \\
\textbf{SEATS} & 7.45\% & {[}2.4\%, 11.7\%{]} & 6.67x {[}15 iter{]} & {[}1.8x, 12.5x{]} \\
\textbf{Twitter} & 4.38\% & {[}1.7\%, 6.8\%{]} & 6.62x {[}13 iter{]} & {[}5.4x, 8.6x{]} \\
\textbf{RS} & 1.08\% & {[}0.2\%, 1.9\%{]} & 1.98x {[}49 iter{]} & {[}1.1x, 3.1x{]} \\
\bottomrule
\end{tabular}
}
\vspace{-3ex}
\end{table}

\noindent\textbf{Workloads.}
For our experiments, we use six \revision{OLTP} workloads with different characteristics (shown in Table~\ref{tab:workload}).
We use the official YCSB suite~\cite{ycsb}, and BenchBase (formerly OLTP-Bench~\cite{oltpbench}).
We use two workloads variants from the YCSB suite: the read-write ($50\%$-$50\%$) balanced YCSB-A, and the read-heavy ($95\%$ read, $5\%$ write), YCSB-B.
YCSB workloads perform simple random access key queries, on a single $10$-column table, based on a Zipfian distribution.
From BenchBase, we use the TPC-C, SEATS, Twitter, and ResourceStresser (RS) workloads.
Both TPC-C and SEATS are traditional OLTP workloads with multiple tables, and complex relations.
TPC-C consists of five (mostly write-heavy) transactions that operate on nine tables, and simulates an order processing system for a warehouse.
SEATS resembles the back-end system of an airline ticketing service using ten tables and six transaction types.
Twitter
is a web-oriented workload that uses public traces to simulate the core functionality of a micro-blogging application; it consists of five tables and five heavily-skewed transaction types.
RS is a synthetic workload with the intention to introduce independent contention on different resources (i.e., CPU, disk I/O and locks)~\cite{oltpbench}.
We employ $40$ clients, and select the scale factor of all workloads accordingly, so that all databases are $20$GB.
\revision{In the future, we also plan to evaluate \sys's set of techniques with OLAP workloads}.

\begin{figure*}[tp]
    \centering
    \begin{minipage}{0.75\textwidth}
    \begin{minipage}{0.31\textwidth}
        \includegraphics[width=\textwidth]{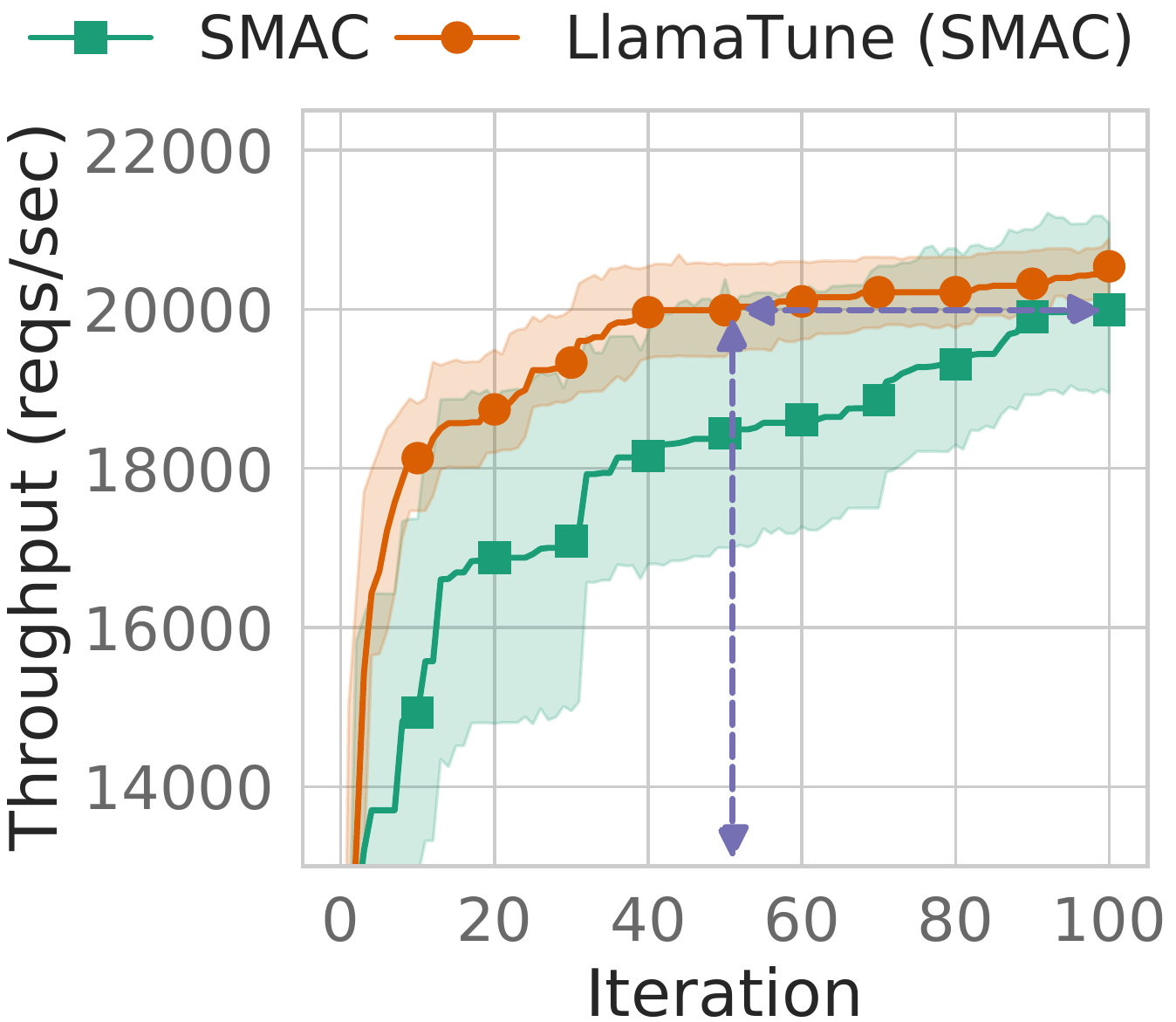}
        \vspace{-0.25in}
        \captionsetup{labelformat=empty}
        \caption{(a) YCSB-A}
    \end{minipage}
    \begin{minipage}{0.33\textwidth}
        \includegraphics[width=\textwidth]{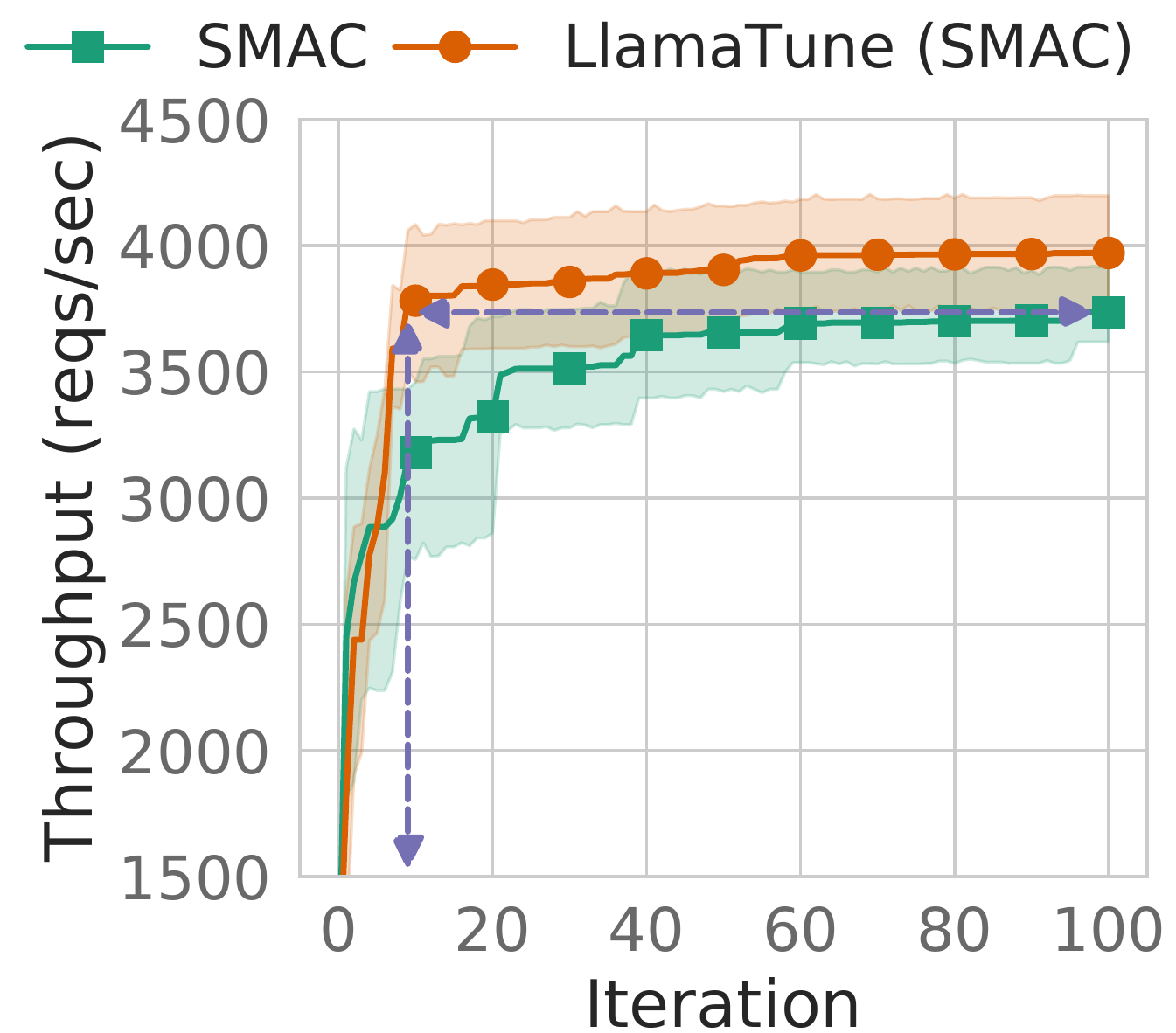}
        \vspace{-0.25in}
        \captionsetup{labelformat=empty}
        \caption{(b) TPC-C}
    \end{minipage}
    \begin{minipage}{0.33\textwidth}
        \includegraphics[width=\textwidth]{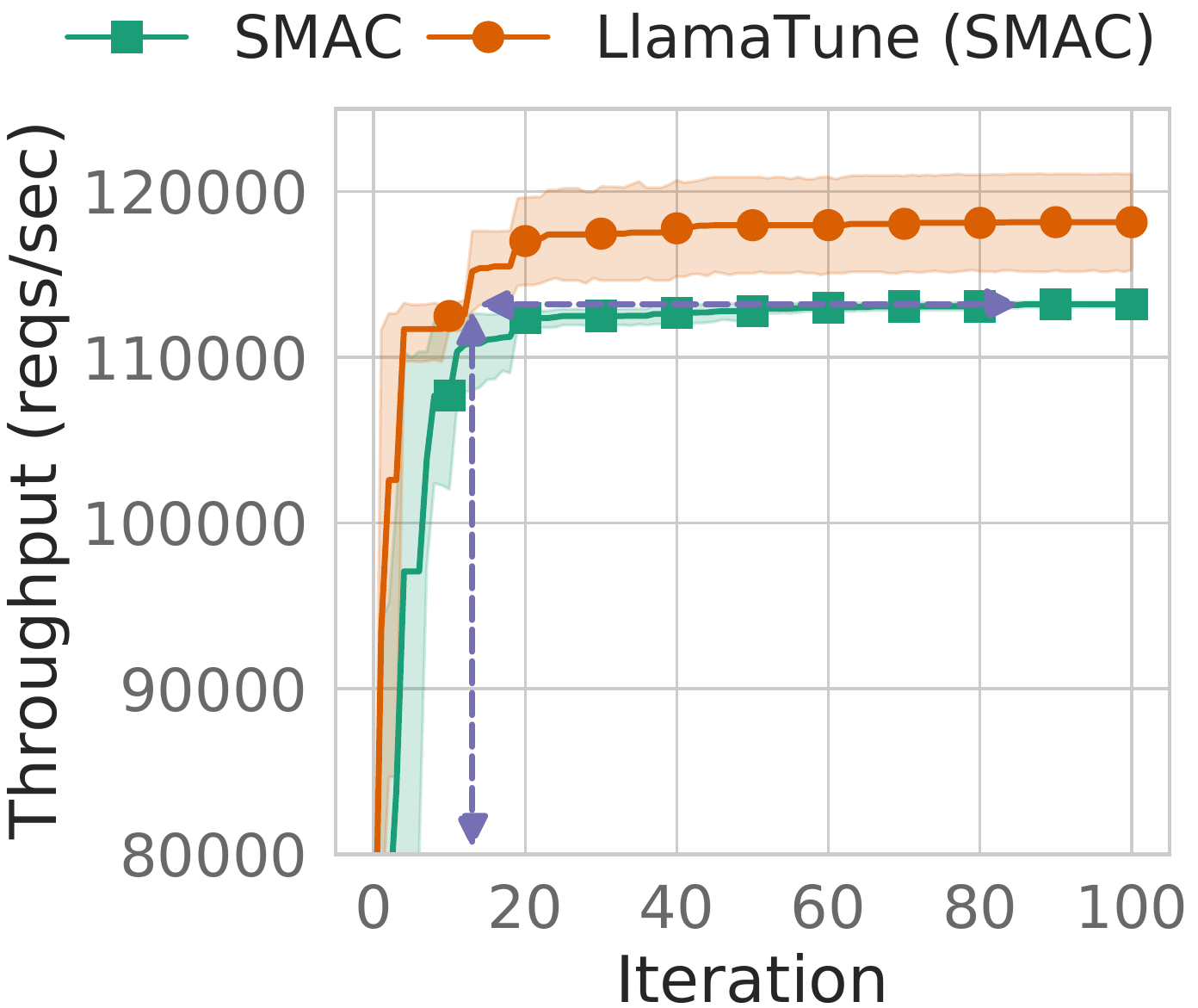}
        \vspace{-0.25in}
        \captionsetup{labelformat=empty}
        \caption{(c) Twitter}
    \end{minipage}
    \addtocounter{figure}{-3}
    \vspace{-0.15in}
    \caption{Best throughput achieved by \sys. Time-to-optimal also shown.\textsuperscript{\ref{note3}}}
    \label{fig:throughput}
    \end{minipage}
    \begin{minipage}{0.21\textwidth}
        \includegraphics[width=\textwidth]{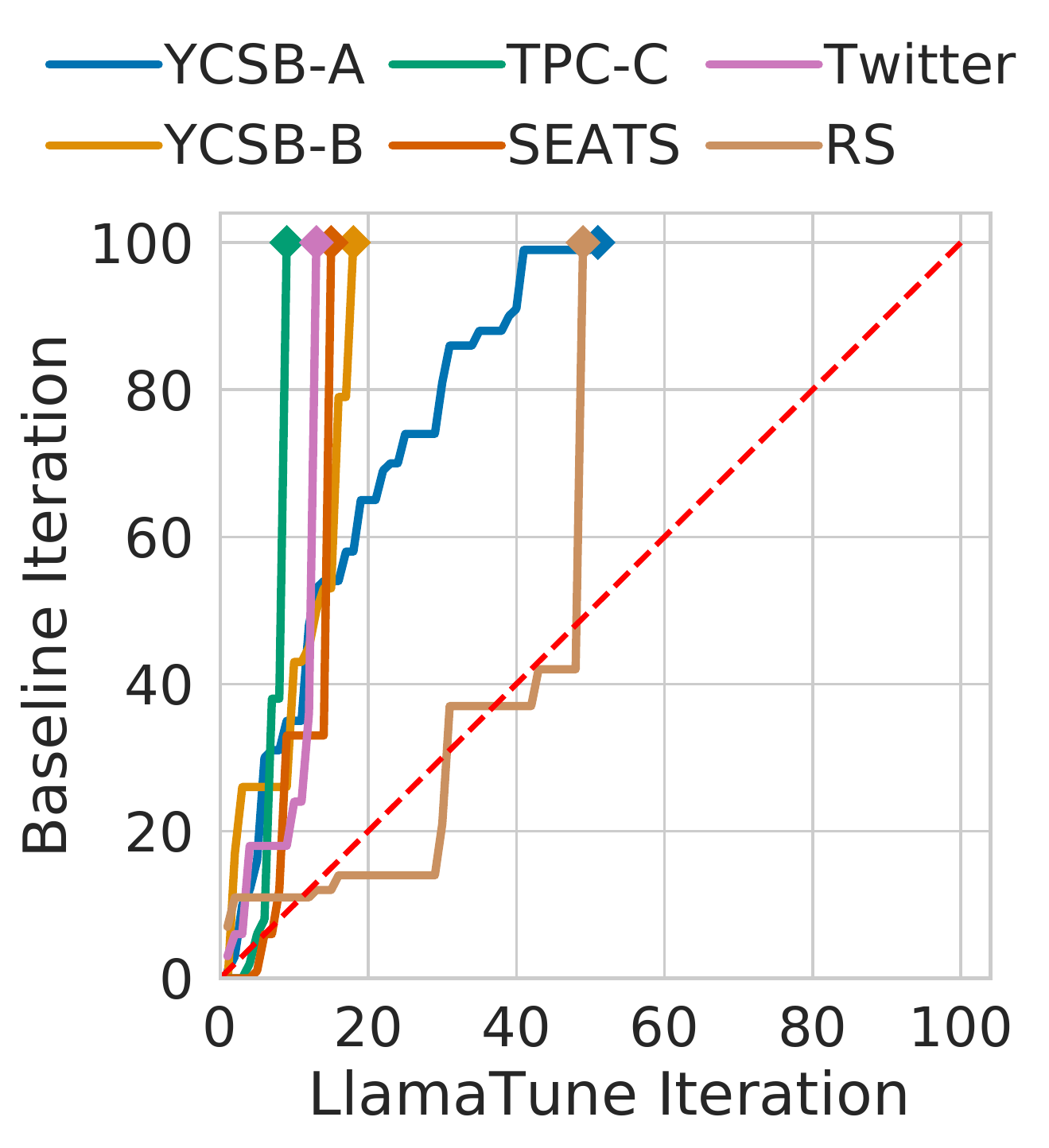}
        \vspace{-0.25in}
        \caption{\revisionenv \sys convergence gains vs. SMAC.}
        \label{fig:llamatune-summary}
    \end{minipage}
    \vspace{-2ex}
\end{figure*}

\noindent\textbf{Optimizers.}
\revision{For our experiments we use three state-of-the-art optimizers: SMAC, GP-BO, and DDPG}.
\revision{All three are shown to outperform alternative methods in most cases~\cite{zhang2021facilitating}.}
For SMAC we use the implementation provided by the authors~\cite{lindauer2021smac3}, while for the GP-BO we leverage OpenBox's~\cite{openbox} implementation, also used in prior work~\cite{zhang2021facilitating}.
\revision{For DDPG, we implement the original neural network architecture used in CDBTune~\cite{rltuning} with PyTorch~\cite{paszke2019pytorch}.}

\noindent\textbf{\sys Implementation.}
Since \sys is agnostic to the underlying optimizer, we implement it in Python3 with $800$ LOCs on top of all three optimizers.
In the experiments, \sys uses HeSBO random projections with $d=16$ dimensions, sets the special value biasing to $20\%$, and bucketizes the search space to limit the number of unique values of each dimension to $K=10,000$.

\noindent\textbf{Evaluation Metrics.}
We use two metrics to evaluate the performance of \sys{}: final DBMS throughput (latency) improvement, and time-to-optimal throughput (latency).
The former is related to the difference in performance (e.g., throughput, $95$th $\%$-tile latency) of the best-performing configurations found at the end of tuning (i.e., after 100 iterations).
We compare the performance improvement (in $\%$) realized by \sys{} compared to the baseline.
The latter is inspired from a popular metric used in ML (i.e., \textit{time-to-accuracy}) that quantifies the time taken by ML algorithms to reach a certain accuracy level~\cite{tta}.
We report the \textit{earliest} iteration at which \sys{} has found a better-performing configuration compared to the baseline optimal, as well as the relative speedup.

\vspace{-1ex}
\subsection{Efficiency Evaluation}

\begin{table}[t]
\caption{Perf. gains of \sys when coupled with SMAC, and tuning for $95$-th percentile tail latency.}
\label{tab:latency}
\vspace{-1em}
\scalebox{0.9}{
\begin{tabular}{ccccc}
\toprule
\small
\multirow{2}{*}{\textbf{Workload}} & \multicolumn{2}{c}{\textbf{\begin{tabular}[c]{@{}l@{}}Final 95th \%-tile \\ Latency Reduction\end{tabular}}} & \multicolumn{2}{c}{\textbf{\begin{tabular}[c]{@{}c@{}}Time-to-Optimal \\ Speedup\end{tabular}}} \\
 & Average & {[}5\%, 95\%{]} CI & Average & {[}5\%, 95\%{]} CI \\
 \midrule
\textbf{TPC-C} & 14.56\% & {[}-1.1\%, 30.8\%{]} & 2.14x {[}43 iter{]} & {[}0.9x, 18.4x{]} \\
\textbf{SEATS} & 11.16\% & {[}1.3\%, 21.2\%{]} & 2.35x {[}34 iter{]} & {[}1.2x, 4.7x{]} \\
\textbf{Twitter} & 3.33\% & {[}-3.9\%, 10.9\%{]} & 1.38x {[}68 iter{]} & {[}0.9x, 3.4x,{]} \\
\bottomrule
\end{tabular}
}
\vspace{-1ex}
\end{table}

\begin{table}[t]
\revisionenv
\caption{Perf. gains of \sys when coupled with SMAC, and tuning the newer DBMS, \mbox{PostgreSQL} v13.6.}
\label{tab:e2e-pg13}
\vspace{-1em}
\scalebox{0.9}{
\begin{tabular}{ccccc}
\toprule
\multirow{2}{*}{\textbf{\small Workload}} & \multicolumn{2}{c}{\textbf{\begin{tabular}[c]{@{}c@{}}Final Throughput \\ Improvement\end{tabular}}} & \multicolumn{2}{c}{\textbf{\begin{tabular}[c]{@{}c@{}}Time-to-Optimal \\ Speedup\end{tabular}}} \\
 & Average & {[}5\%, 95\%{]} CI & Average & {[}5\%, 95\%{]} CI \\
 \midrule
\textbf{YCSB-A} & 10.73\% & {[}8.9\%, 13.0\%{]} & 2.69x {[}36 iter{]} & {[}2.1x, 7.5x{]} \\
\textbf{YCSB-B} & 3.63\% & {[}2.3\%, 4.9\%{]} & 2.54x {[}37 iter{]} & {[}1.4x, 3.9x{]} \\
\textbf{TPC-C} & 2.91\% & {[}1.1\%, 4.2\%{]} & 1.56x {[}63 iter{]} & {[}1.4x, 4.3x{]} \\
\textbf{SEATS} & 20.25\% & {[}16.7\%, 23.5\%{]} & 6.0x {[}16 iter{]} & {[}3.4x, 16.0x{]} \\
\textbf{Twitter} & 2.61\% & {[}0.8\%, 4.3\%{]} & 8.25x {[}12 iter{]} & {[}5.2x, 12.4x{]} \\
\textbf{RS} & 0.69\% & {[}0.0\%, 1.4\%{]} & 2.12x {[}42 iter{]} & {[}0.9x, 4.1x{]} \\
\bottomrule
\end{tabular}
}
\vspace{-2ex}
\end{table}

In this section, we evaluate the tuning efficiency of \sys while using the vanilla SMAC optimizer as our baseline.

\noindent\textbf{Optimizing for Throughput.}
\label{sec:end-to-end-throughput}
We set the optimizer target to throughput; Table~\ref{tab:throughput-smac} shows the results for all six workloads.
We make two observations.
First, \sys reaches the performance of the baseline SMAC optimum configuration by $\sim5.62\times$ faster on average.
For four of the total six workloads, it reaches that performance before iteration $20$, which highlights its increased sample-efficiency.
Second, \sys can improve the average final throughput (after $100$ iterations) on all workloads, by $\sim7.13\%$ on average, compared to vanilla SMAC (from $1.08\%$ better on RS to $20.85\%$ better on YCSB-B).
Both TPC-C, SEATS, which are complex OLTP workloads show consistent gains of $6$-$7\%$, while Twitter which has vastly different characteristics, also performs well.
The synthetic RS workload is particularly hard to optimize, as the overall performance gains compared to the default configuration is only around $10\%$; this explains why \sys's improvement is small, compared to the other workloads.

Figure~\ref{fig:throughput} shows the best performance convergence curve for three workloads: YCSB-A, TPC-C, and Twitter.
The red line indicates the earliest iteration of \sys that results in better performance compared to the baseline (i.e., time-to-optimal). 
The contributing factors for the improvement include projection to a low-dimensional space, as well as special value biasing, which accelerates exploring their effect on performance.
We also observe that for YCSB-A, \sys reduces the variance across different runs, making it more likely for any run to realize the highest gains.

{
\revisionenv
Figure~\ref{fig:llamatune-summary} shows the mean improvement of \sys over the baseline (SMAC) during the entire tuning session.
For every iteration of \sys (x-axis), we plot the \textit{earliest} iteration at which SMAC achieves the same (or better) performance on the (y-axis); the point at which \sys surpasses the best SMAC performance is indicated by a diamond. We see that for all workloads other than RS, \sys{} provides improvements over SMAC across all iterations. For RS, we see that \sys{} performs better from around iteration 30. Overall, we can see that given any fixed iteration budget, \sys{} can provide meaningful speedups across workloads.
}

\vspace{1ex}
\noindent\textbf{Optimizing for Tail Latency.}
Now, we experiment with a different tuning scenario, where we aim to reduce the $95$-th percentile latency, at a fixed rate of incoming requests.
We experiment with three workloads from the BenchBase suite: TPC-C, SEATS, and Twitter.
We set the rate of incoming requests to half of the best throughput achieved from the previous experiments: $2,000$ requests per second for TPC-C, $8,000$ for SEATS, and $60,000$ for Twitter.
Table~\ref{tab:latency} summarizes the results.
We observe that \sys outperforms the baseline SMAC on all three workloads.
On average \sys yields a $\sim1.96\times$ time-to-optimal speedup, 
and $\sim9.68\%$ better final tail latency.
These results indicate that our tuning pipeline is still effective in more complicated tuning targets.

\begin{figure*}[tp]
    \begin{minipage}{0.55\textwidth}
        \includegraphics[width=0.69\textwidth, trim={0 4.7in 0 0},clip]{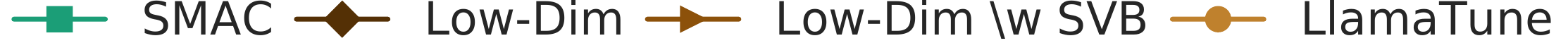}
        \begin{minipage}{0.32\textwidth}
            \includegraphics[width=\textwidth]{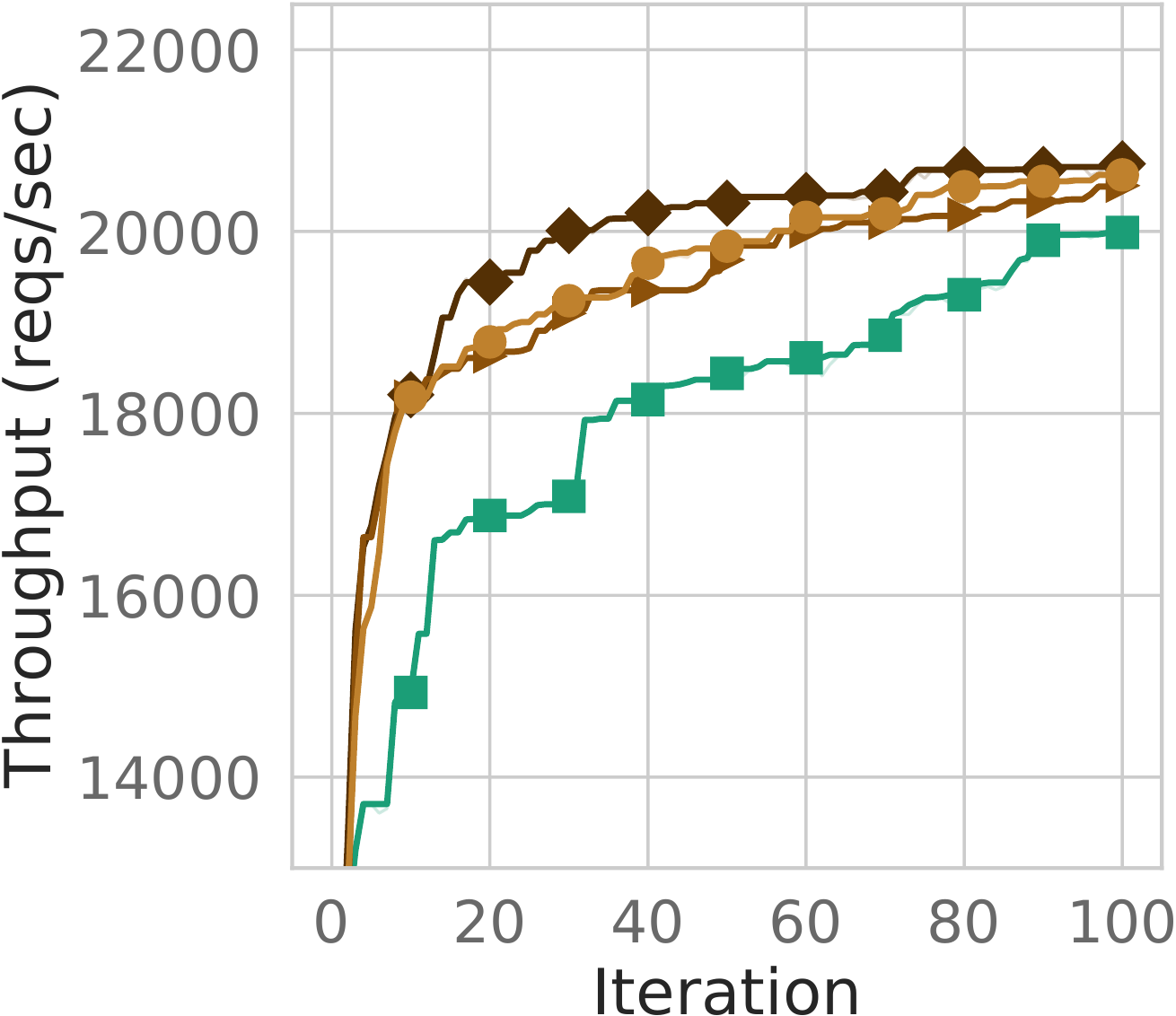}
            \vspace{-0.25in}
            \captionsetup{labelformat=empty}
            \caption{(a) YCSB-A}
        \end{minipage}
        \begin{minipage}{0.33\textwidth}
            \includegraphics[width=\textwidth]{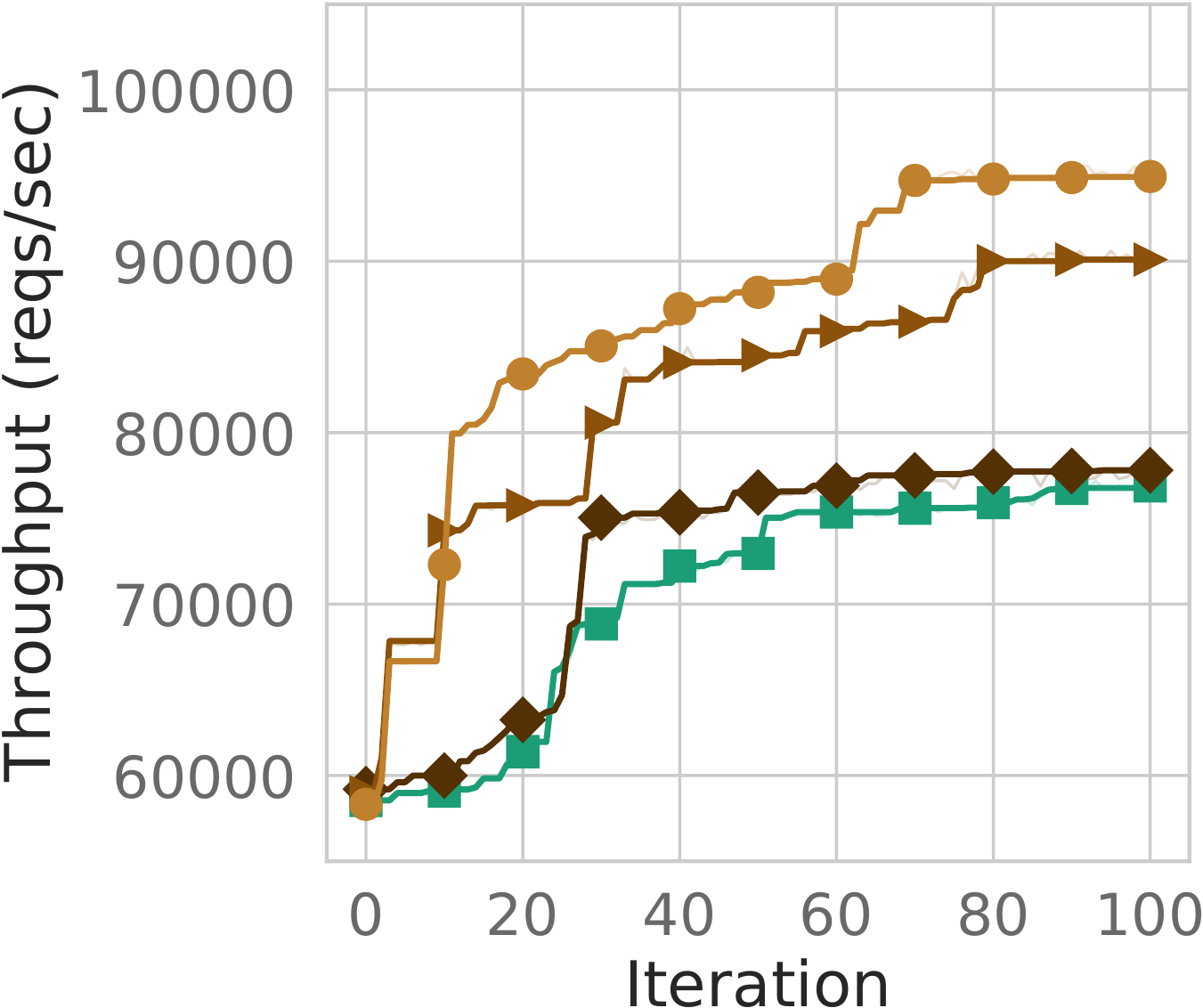}
            \vspace{-0.25in}
            \captionsetup{labelformat=empty}
            \caption{(b) YCSB-B}
        \end{minipage}
        \begin{minipage}{0.31\textwidth}
            \includegraphics[width=\textwidth]{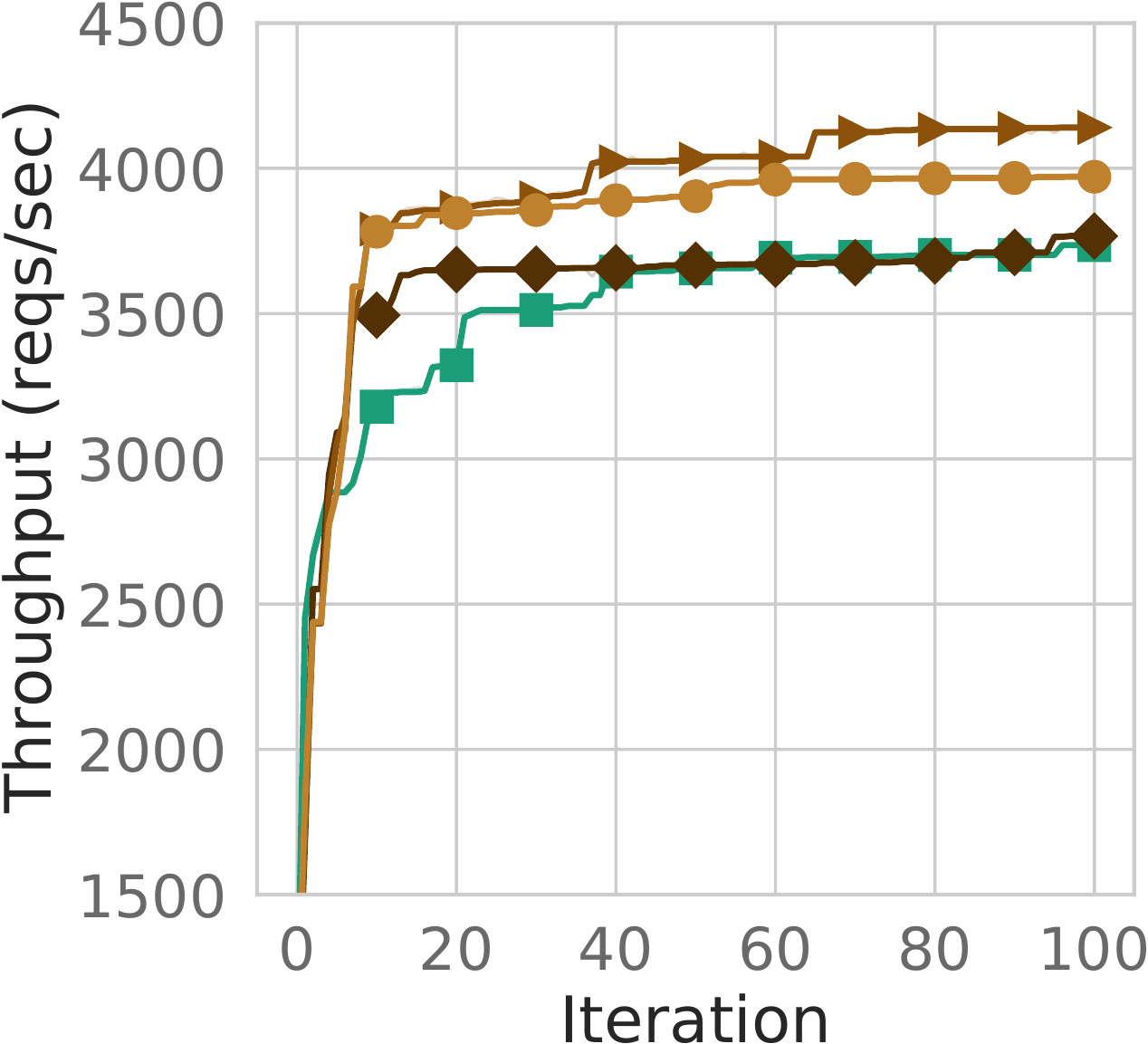}
            \vspace{-0.25in}
            \captionsetup{labelformat=empty}
            \caption{(c) TPC-C}
        \end{minipage}
        \addtocounter{figure}{-3}
        \vspace{-0.13in}
        \caption{Ablation study for the three components of \sys{}.\textsuperscript{\ref{note3}}}
        \label{fig:ablation}
    \end{minipage}
    \hspace{.5em}
    \begin{minipage}{0.42\textwidth}
        \vspace{-1ex}
        \captionof{table}{Performance gains of \sys when \mbox{coupled} with a different BO-based method, GP-BO.}
        \label{tab:throughput_gpbo}
        \vspace{-1em}
        \scalebox{0.8}{
        \begin{tabular}{ccccc}
        \toprule
        \multirow{2}{*}{\textbf{Workload}} & \multicolumn{2}{c}{\textbf{\begin{tabular}[c]{@{}c@{}}Final Throughput \\ Improvement\end{tabular}}} & \multicolumn{2}{c}{\textbf{\begin{tabular}[c]{@{}c@{}}Time-to-Optimal \\ Speedup\end{tabular}}} \\
         & Average & {[}5\%, 95\%{]} CI & Average & {[}5\%, 95\%{]} CI \\
        \midrule
        \textbf{YCSB-A} & 1.10\% & {[}-3.1\%, 6.1\%{]} & 1.11x {[}90 iter{]} & {[}0.8x, 3.1x{]} \\
        \textbf{YCSB-B} & 21.54\% & {[}10.9\%, 33.3\%{]} & 19.4x {[}5 iter{]} & {[}8.8x, 32.3x{]} \\
        \textbf{TPC-C} & 18.57\% & {[}10.8\%, 26.3\%{]} & 10.38x {[}8 iter{]} & {[}4.9x, 16.6x{]} \\
        \textbf{SEATS} & 10.06\% & {[}8.1\%, 11.9\%{]} & 3.5x {[}28 iter{]} & {[}1.3x, 19.6x{]} \\
        \textbf{Twitter} & 4.28\% & {[}1.4\%, 7.1\%{]} & 14.83x {[}6 iter{]} & {[}3.4x, 17.8x{]} \\
        \textbf{RS} & 0.50\% & {[}-0.9\%, 1.6\%{]} & 1.43x {[}70 iter{]} & {[}0.8x, 7.1x{]} \\
        \bottomrule
        \end{tabular}
        }
    \end{minipage}
    \vspace{-2ex}
\end{figure*}

{
\revisionenv
\vspace{-2ex}
\subsection{Effectiveness on Different DBMS Version}
\label{subsec:eval-pg13}

We next experiment with a newer version of PostgreSQL, v13.6, to evaluate \sys's effectiveness and robustness when the underlying DBMS characteristics change.
Compared to the version used throughout the paper (v9.6), this newer version includes several major improvements from just-in-time compilation to better parallel query execution; this is also reflected in the increased number of configuration knobs~\cite{postgresql-documentation}.
Porting \sys to the new version requires characterizing which of the newly-introduced knobs should be included in the set of tunable knobs, and identifying new hybrid knobs (along with their special value).
Overall it took us $\sim 4$ hours to fully integrate \sys with this version.
We now tune $112$ knobs ($23$ of which are hybrid ones) using the \textit{same hyperparameters} for \sys, as before, and we compare against vanilla SMAC; the rest of the setup is identical to section~\ref{sec:end-to-end-throughput}.
Table~\ref{tab:e2e-pg13} shows the results.
On average, \sys achieves $\sim3.86\times$ time-to-optimal speedup, and manages to match or outperform the vanilla SMAC in all cases.
Interestingly, we also observe that the recent PostgreSQL improvements affect the gains realized by \sys. For example, with YCSB-B, while \sys{} achieves similar throughput across Postgres versions, we find that SMAC achieves better throughput with v13.6, narrowing the throughput  improvement of \sys over SMAC from 20.8\% to 3.6\%.
On the other hand, we see better throughput improvements with \sys{} for SEATS and YCSB-A.
}

\vspace{-2ex}
\subsection{Generalizing across Optimizers}
\label{subsec:eval-generalization-optimizer}

\noindent
\textbf{GP-BO.}
We now change the underlying optimizer of \sys to GP-BO, which uses Gaussian Processes as surrogate model, instead of SMAC's random forests.
We conduct the same set of experiments as section~\ref{sec:end-to-end-throughput}, optimizing for maximum throughput.
Table~\ref{tab:throughput_gpbo} shows the results.
We observe that \sys again outperforms the vanilla GP-BO optimizer on all six workloads.
Across all workloads, the mean time-to-optimal speedup is $\sim8.4\times$, while \sys also finds better-performing configurations.
In particular, we see much faster convergence-to-baseline optimal for YCSB-B ($19.4\times$), TPC-C ($10.38\times$), and Twitter ($14.83\times$) where less than $10$ iterations are required.
RS still remains a challenging workload to optimize, as \sys could not improve much over the baseline GP-BO.

\begin{table}[t]
\revisionenv
\caption{\revisionenv Perf. gains of \sys when coupled with DDPG.}
\label{tab:ddpg}
\vspace{-1em}
\scalebox{0.86}{
\begin{tabular}{lcccc}
\toprule
\multirow{2}{*}{\textbf{\small Workload}} & \multicolumn{2}{c}{\textbf{\begin{tabular}[c]{@{}c@{}}Final Throughput \\ Improvement\end{tabular}}} & \multicolumn{2}{c}{\textbf{\begin{tabular}[c]{@{}c@{}}Time-to-Optimal\\ Speedup\end{tabular}}} \\
 & Average & {[}5\%, 95\%{]} CI & Average & {[}5\%, 95\%{]} CI \\
\midrule
\textbf{YCSB-B} & 24.95\% & {[}15.85\%, 32.69\%{]} & 5.17x {[}18 iter{]} & {[}2.38x, 6.64x{]} \\
\textbf{TPC-C} & 12.24\% & {[}-1.77\%, 30.27\%{]} & 8.40x {[}~5 iter{]} & {[}0.82x, 21.0x{]} \\
\textbf{Twitter} & 2.34\% & {[}-0.60\%, 5.14\%{]} & 1.18x {[}66 iter{]} & {[}0.75x, 13.0x{]} \\
\textbf{RS} & 0.71\% & {[}-1.16\%, 2.47\%{]} & 1.79x {[}34 iter{]} & {[}0.74x, 3.39x{]} \\
\bottomrule
\end{tabular}
}
\vspace{-3ex}
\end{table}

{
\revisionenv
\noindent
\textbf{DDPG.}
We next evaluate \sys when it is coupled with DDPG, a reinforcement learning-based optimizer (used in CDBTune~\cite{rltuning} and QTune~\cite{qtune}).
DDPG leverages two neural networks: the \textit{actor} that recommends a configuration when given the internal DBMS state (metrics) as input, and the \textit{critic} that assesses how good each configuration is, based on its resulting performance (reward).
We use a similar setup as in ~\ref{sec:end-to-end-throughput} and maximize throughput.
To represent the internal DBMS state, we sample $27$ system-wide PostgreSQL metrics every $5$ seconds, and feed their average into the neural network at the end of each iteration.
Table~\ref{tab:ddpg} presents the results for four workloads.
On average, \sys achieves a time-to-optimal speedup of $4.14\times$, while also finding better-performing configurations (up to $25\%$ for YCSB-B).
Thus, overall we can see that our benefits extend across optimizers.
}

\vspace{-1ex}
\subsection{Ablation Study}
\label{sec:ablation-study}

We next conduct an ablation study to better understand the benefits from each of \sys's components.
We compare using only the HeSBO-16 low-dimensional projection, with HeSBO-16 plus special value biasing (SVB), and with the entire \sys pipeline (i.e. with configuration space bucketization).
We compare \sys{} to SMAC on three workloads: YCSB-A, YCSB-B, and TPC-C (Figure~\ref{fig:ablation}). 
We observe that all three combinations perform as well or better compared to the SMAC baseline, for all three workloads. 
For YCSB-B we see that only using HeSBO-16 achieves $\sim2\times$ time-to-optimal speedup, while using the other techniques this speedup is improved to $\sim5.5\times$.
For YCSB-A, the plain HeSBO-16 projection performs the best, while it seems that the special value biasing hinders \sys's convergence; the final throughput is not affected though.
This is because special knobs values do not improve the YCSB-A performance, and thus biasing the optimizer sampling towards those, slightly delays exploring other better-performing regions.
For TPC-C, we observe that HeSBO-16 and special value biasing have a positive effect on optimization performance.
We note a small performance degradation when applying the search space bucketization and believe that this is because of \sys imposing a bucketized space for \textit{all} knobs, not just for the ones with a large number of unique values.
Finally, we note that even though the integration of special value biasing and space bucketization do not always help, their (negative) impact on performance is very limited; when they do, we can realize significant gains.

\vspace{-1ex}
\subsection{Optimizer Overhead}

\begin{table}[t]
\caption{Optimizer overhead \& \sys improvements}
\label{tab:opt_overhead}
\vspace{-1em}
\scalebox{0.95}{
\begin{tabular}{cccc}
\toprule
\multirow{2}{*}{\textbf{Optimizer}} & \multicolumn{2}{c}{\textbf{Time Overhead (minutes)}} & \multirow{2}{*}{\textbf{\begin{tabular}[c]{@{}c@{}}Time \\ Reduction (\%)\end{tabular}}} \\ \cline{2-3}
& Baseline & LlamaTune & \\ \hline
\textbf{SMAC} & 26.75 & 3.83 & 86 \\
\textbf{GP-BO} & 256.57 & 62.95 & 75 \\
\textbf{\revision{DDPG}} & \revision{0.075} & \revision{0.066} & \revision{12} \\
\bottomrule
\end{tabular}
}
\vspace{-3ex}
\end{table}

We measure the time taken by the optimizer to suggest the next configuration across the entire tuning session (i.e., $100$ iterations).
This time includes updating the underlying surrogate models, as well as comparing the set of possible configuration candidates; it does \textit{not} include the time for a configuration to be evaluated on the DBMS.
Table~\ref{tab:opt_overhead} summarizes the time overhead of \sys compared to the vanilla optimizers.
\sys reduces the time overhead by $86\%$, $75\%$, $12\%$ when being integrated with SMAC, GP-BO, and DDPG, respectively.
The reduced number of dimensions, resulting from the low-dimensional projection is the primary reason, as the optimizers have to model a much smaller space.
This is especially important for the GP-BO optimizer, due to the non-linear sampling behavior of GPs~\cite{shahriari2015taking}.
The time overhead of special values biasing, and knob values bucketization are negligible.

\vspace{-1ex}
\section{Conclusion}

In this paper, we studied techniques to improve the sample efficiency of optimizers used in DBMS tuning and 
showed how we can leverage domain knowledge to perform low dimensional tuning while handling special knob values and large knob value ranges. 
We empirically showed that our techniques can find good performing configurations up to $11\times$ faster compared to the state-of-the-art optimizer (SMAC), and that our tuning pipeline is robust when tuning for different performance targets, optimizers, and DBMS versions.
In the future, we plan to extend our work to other contexts, such as tuning Linux or dataflow engines like Apache Spark.

\vspace{-1ex}
\section*{Acknowledgements}
We thank the anonymous reviewers for their insightful comments that improved this paper. This work was supported in part by a grant from Microsoft Gray Systems Lab and by the Office of the Vice Chancellor for Research and Graduate Education at UW-Madison with funding from the Wisconsin Alumni Research Foundation.

\bibliographystyle{ACM-Reference-Format}
\bibliography{papers,urls}

\begin{appendix}
    \noindent\textbf{\LARGE APPENDIX}

\section{Deployment Scenario}

\begin{table}
    \captionof{table}{Evaluation for three early-stopping \mbox{policies}. We report mean performance improvement and the iteration when the session was stopped.}
    \label{tab:early-stop}
    \vspace{-1em}
    \scalebox{1}{
    \begin{tabular}{c|cc|cc|cc}
    \toprule
    \textbf{Policies} & \multicolumn{2}{c|}{\textbf{(0.5\%, 10)}} & \multicolumn{2}{c|}{\textbf{(1\%, 10)}} & \multicolumn{2}{c}{\textbf{(1\%, 20)}} \\ \cline{1-7}
     \textbf{Metrics} & \begin{tabular}[c]{@{}c@{}}Perf. \\ Improv \\ (\%)\end{tabular} & \begin{tabular}[c]{@{}c@{}}Num \\ Iters\end{tabular} & \begin{tabular}[c]{@{}c@{}}Perf. \\ Improv \\ (\%)\end{tabular} & \begin{tabular}[c]{@{}c@{}}Num\\ Iters\end{tabular} & \begin{tabular}[c]{@{}c@{}}Perf. \\ Improv \\ (\%)\end{tabular} & \begin{tabular}[c]{@{}c@{}}Num\\ Iters\end{tabular} \\
    \midrule
    \textbf{YCSB-A} & -0.01 & 49 & -0.01 & 46 & 1.12 & 71 \\
    \textbf{YCSB-B} & 11.82 & 42 & 11.67 & 35 & 20.85 & 96 \\
    \textbf{TPC-C} & 3.08 & 26 & 3.08 & 26 & 6.11 & 72 \\
    \textbf{SEATS} & 1.82 & 31 & 1.55 & 25 & 6.71 & 76 \\
    \textbf{Twitter} & 3.76 & 29 & 3.76 & 29 & 3.82 & 39 \\
    \textbf{RS} & -1.71 & 26 & -2.82 & 12 & 0.82 & 69 \\
    \bottomrule
    \end{tabular}
    }
\end{table}

We consider a deployment scenario in which the user wishes to tune a database for a new workload. 
Our goal here is to study how the gains in terms of convergence speed (i.e. time-to-optimal) can be realized in this setting.
Specifically, we stop the tuning session \textit{early}, once we believe that \sys has found a good configuration, thus saving valuable time and resources.
Ideally, the DBMS performance at this point will be equal or better than running the vanilla optimizer for a fixed time budget (e.g., 100 iterations).

To this end, we augment \sys with an early-stopping policy used in the ML community~\cite{early-stopping}.
This policy continuously monitors the best performance achieved at each iteration, and makes a decision on when tuning should be stopped.
Formally, it can be expressed by two parameters: the minimum best performance improvement ($x$ in $\%$), and the maximum number of $k$ iterations to wait for this improvement to be made (patience).
If $k$ iterations have passed and the aggregate performance improvement is \textit{smaller} than the one required ($x$), then the session is terminated.

We experiment with three (min-improv, patience) policy configurations on all six workloads.
Table~\ref{tab:early-stop} shows the final performance improvement over the vanilla SMAC when tuning stops.
We first focus on the $(1\%, 10)$ early-stopping policy configuration.
When using the $(1\%, 10)$ early-stopping policy, we see that for most workloads \sys achieves better (up to $11.67\%$ for TPC-C) or equal final performance, compared to the SMAC baseline, while stopping after 29 iterations on average.
For RS, the policy terminates very early ($12$th iteration), which results in slight decrease in final throughput.
We can alleviate this behavior by choosing a more conservative policy.
For instance, reducing the minimum improvement threshold to $0.5\%$ (i.e., left column) results in slight increase in the final throughput for YCSB-B, SEATS, and most notably RS, as the policy makes the decision to stop around $10$ iterations later in the tuning process.
Conversely, we can increase the policy patience to $20$ iterations (i.e., right column) to realize near-optimal gains for all workloads, at the expense of spending around $70$ iterations to tune on average.
We allow users to adjust the early-stopping policy in order to trade-off between better final performance vs. earliest stopping possible.
The above results demonstrate that it is possible to realize \sys's gains for tuning new DBMS workload deployments.

\end{appendix}

\end{document}